\documentclass{article}
\usepackage[numbers]{natbib}

\usepackage{arxiv}
\usepackage{epstopdf}
\usepackage[utf8]{inputenc} 
\usepackage[T1]{fontenc}    
\usepackage{hyperref}       
\usepackage{url}            
\usepackage{booktabs}       
\usepackage{amsfonts}       
\usepackage{nicefrac}       
\usepackage{microtype}      
\usepackage{lipsum}		
\usepackage{graphicx}
\usepackage{natbib}
\usepackage{doi}
\usepackage{multirow}
\usepackage{array}
\usepackage{hyperref}       

\AtBeginDocument{%
  \providecommand\BibTeX{{%
    \normalfont B\kern-0.5em{\scshape i\kern-0.25em b}\kern-0.8em\TeX}}}

\title{Blockchain-empowered Data-driven Networks: A Survey and Outlook}

\author{ {\hspace{1mm}Xi Li}\\
	\texttt{lixi@bupt.edu.cn} \\
	\And
	{\hspace{1mm}Zehua Wang} \\
	\texttt{zwang@ece.ubc.ca} \\
	\And
	{\hspace{1mm}Victor C.M. Leung} \\
	\texttt{vleung@ieee.org} \\
	\And
	{\hspace{1mm}Hong Ji} \\
	\texttt{jihong@bupt.edu.cn} \\
	\And
	{\hspace{1mm}Yiming Liu} \\
	\texttt{liuyiming@bupt.edu.cn} \\
	\And
	{\hspace{1mm}Heli Zhang} \\
	\texttt{zhangheli@bupt.edu.cn} \\
}

\renewcommand{\shorttitle}{\textit{arXiv} Template}

\begin{document}
\maketitle
\begin{abstract}

The paths leading to future networks are pointing towards a data-driven paradigm to better cater to the explosive growth of mobile services as well as the increasing heterogeneity of mobile devices, many of which generate and consume large volumes and variety of data. These paths are also hampered by significant challenges in terms of security, privacy, services provisioning, and network management.
Blockchain, which is a technology for building distributed ledgers that provide an immutable log of transactions recorded in a distributed network, has become prominent recently as the underlying technology of cryptocurrencies and is revolutionizing data storage and processing in computer network systems.
For future data-driven networks (DDNs), blockchain is considered as a promising solution to enable the secure storage, sharing, and analytics of data, privacy protection for users, robust, trustworthy network control, and decentralized routing and resource managements.
However, many important challenges and open issues remain to be addressed before blockchain can be deployed widely to enable future DDNs.
In this article, we present a survey on the existing research works on the application of blockchain technologies in computer networks, and identify challenges and potential solutions in the applications of blockchains in future DDNs. We identify application scenarios in which  future blockchain-empowered DDNs could improve the efficiency and security, and generally the effectiveness of network services.

\end{abstract}

%

\maketitle

\section{Introduction}

With the explosive increase of network users and their myriads of network-connected devices generating and consuming massive amounts of digital contents, future networks will encounter significant challenges to provide intelligent services in an efficient and secured manner to users due to the large volumes of data generated, transmitted and processed in the networks.
According to the 2019 Cisco report~\cite{cisco2018cisco}, global Internet data will increase nearly three-folds in the next five years, and global data traffic will reach 4.8 ZB/year by 2022.
Obviously, the paths towards future networks are pointing towards a data-driven paradigm, {\it i.e.}, using data-driven networks (DDNs)~\cite{fang2019data}~\cite{ITU}, to enhance existing services and enable new services to both humans and devices. However, there are obstacles along these paths arising from challenges including security, privacy, services provisioning, and network management. Among the possible solutions for future DDNs,  blockchain, which gains prominence as the technology for cryptocurrencies, has drawn much attention from both academia and industry and is now considered as a promising direction to address many challenges.
Compared with traditional data tools such as database, blockchain gets rid of the centralized authority, and establishes integrity and transparency in distributed nodes.
Therefore, future blockchain-empowered data-driven networks (BDNs) can provide intelligent services with better quality of  experience (QoE) for users on the premise of more secure data storage, sharing, and analytics, better protection of users' privacy, more robust and trustworthy network control, and decentralized routing and resource managements. Therefore, although research on BDNs is still in the early stage, it is timely and significant to survey the underlying techniques and technologies and shine the light on critical open issues.

Blockchain is a distributed and practically immutable ledger of transactions.
Each block in a blockchain contains one or more transactions and points to the prior one~\cite{ref2}.
As the primary purpose of a blockchain, transaction records can be securely saved and validated in an untrusted peer-to-peer system by applying a consensus mechanism in a decentralized manner.
The consensus mechanism is used to guarantee that the majority of peers can reach an agreement on the transaction records disseminated over the network with gossip protocols.
In addition, transactions can implement and execute the operational codes saved in a blockchain, enabling software services among untrusted users.
Since first used in cryptocurrency~\cite{ref3}, blockchain has attracted increasing interests with lots of use cases and applications.
Particularly, blockchain has been envisioned as a promising solution to many problems when new technologies, including cloud and edge computing~\cite{ref7,ref8}, big data~\cite{ref6}, and Internet of Things (IoT)~\cite{ref4,ref5}, are integrated with networks.

By leveraging blockchain, current DDNs can evolve from many perspectives in the near future.
Future DDNs should be able to provide feasible solutions for dynamic access control, guarantee the integrity and validity of the exchanged data, and preserve the privacy of mobile users.
Blockchains may also be integrated in future DDNs for maintaining tight synchronization among different network elements equipped with storage, computing, and networking resources.
While the advantages are quite inspiring, several significant research challenges need to be well investigated before the widespread implementation and deployment of BDNs will be possible, such as scalability in transaction throughput, decentralized intelligent network control, and resource management and allocation.
However, to the best of our knowledge, these challenges have not been well addressed in previous work. In this survey, we explore the related pioneer research works and investigate the applications of blockchain technology in computer networks to gleam their potential benefits when applied to future DDNs.
A number of research challenges are also identified for BDNs when considering the unique  characteristics of blockchain technology.
Our work takes the first steps toward a better understanding of how blockchains may be embedded in future DDNs to improve the performance of data-driven applications and network management, and opens up pathways to the development and deployment of future BDNs.

\begin{figure*}[!t]
	\centering
	\includegraphics[width=1.0\textwidth]{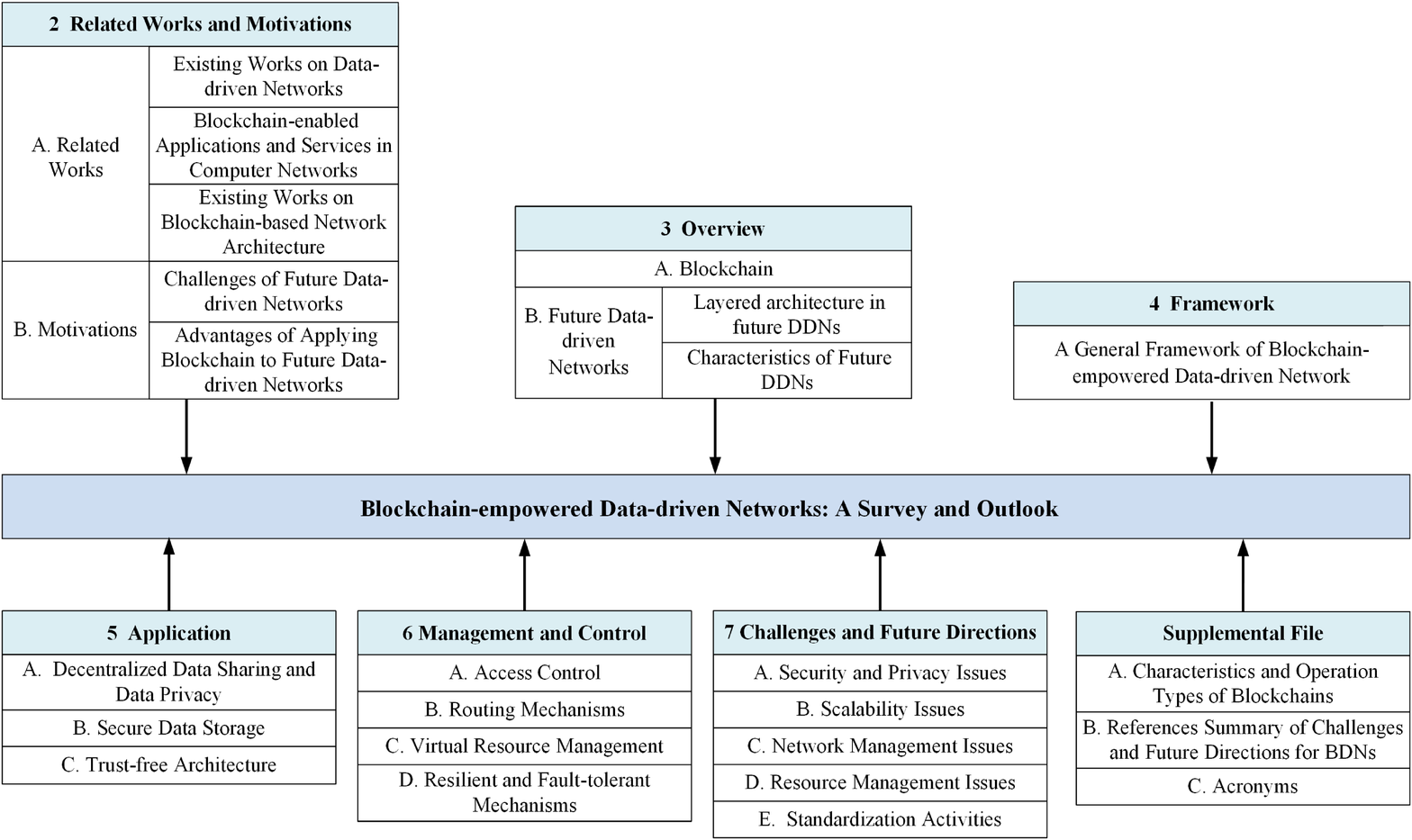}
	\caption{Skeleton of the survey.}
	\label{fig:outline}
	\vspace{-0.2in}
\end{figure*}

To better present our work, we summarize the structure of this article in Fig.~\ref{fig:outline}.
Section~\ref{sec:rltd_wrk_mtv} presents the related works and motivations.
Section~\ref{sec:overview} presents the background of blockchain technology and describes the future DDNs.
Section~\ref{sec:frmwk_of_blkchn_ddn} discusses the framework of future BDNs.
Section~\ref{sec:app_survey} surveys in more details the related works on promising services enabled by future BDNs.
Section~\ref{sec:manage_survey} summarizes the existing works on management and control for future BDNs.
Technical challenges and future directions of BDNs are discussed in Section~\ref{sec:challenges}.
Section~\ref{sec:conclusion} concludes this article.
Please also refer to the online supplemental file for additional materials  including the characteristics and operation types of blockchain, a  summary of references in Section~\ref{sec:challenges}, and a list of acronyms.

\section{Related Works and Motivations}
\label{sec:rltd_wrk_mtv}
\subsection{Related Works}

\subsubsection{Existing Works on Data-driven Networks}

The emergence of big data technologies has sped up the emergence of data-driven Internet to support new applications and efficient allocation of resources~\cite{yin2014big}.
The DDNs can extract valuable information of the network through analysis of the massive amount of data collected over network nodes and end-points, based on which optimal strategies are learned (most likely using machine intelligence that can improve its learning outcome also through big data analytics) and fed back to the network nodes and end-points to enhance network operation and management~\cite{yao2016novel}.
In this subsection, we review some existing works on DDNs, as summarized in Table~\ref{table:ExistingDDN}. These works have fueled the growing interest in DDNs and continued evolution of the concepts of DDNs to leverage growing types of data. Challenges arising from the development toward future DDNs are discussed with more details in Section 2.2.1.

\newcommand{\tabincell}[2]{\begin{tabular}{@{}#1@{}}#2\end{tabular}}
    \begin{table*}[!th]
      \centering
      \caption{Existing research works on Data-driven Networks.}
      \label{table:ExistingDDN}
      \begin{tabular}{m{3.3cm}<{\centering}|m{0.8cm}<{\centering} m{9cm}}
      \hline Focus  &  Ref.  &
      \multicolumn{1}{c}{Contributions}
      \\ \hline

      \multirow{6}{*}{Network architecture}
      &  \cite{wang2016data} & {Propose a data-based network architecture to enhance personalized QoE in 5G networks. }
      \\ \cline{2-3}
      & \cite{yin2014big}  & {Propose a network architecture that collects data and optimizes network from bottom up.}
      \\ \cline{2-3}
      & \cite{han2017big}  & {Propose a mobile network framework enabled by big data to provide efficient resource allocation, content delivery, and RAN optimization services. }
      \\ \cline{2-3}
       & \cite{cui2016big}  & {Discuss the benefits of introducing big data to SDNs. }
       \\   \hline

      \multirow{6}{*}{Resource allocation}
      & \cite{zheng2016big}  & {Introduce a big data-driven mobile network framework to provide best QoE and minimize the cost.}
      \\ \cline{2-3}
      & \cite{fang2019data}  & {Combine ICNs, SDNs and big data to achieve optimal resource allocation and intelligent content delivery.}
      \\ \cline{2-3}
       &  \cite{chen2018data}  & {For computing and caching scenario, design a data-driven  architecture to achieve ultra-low latency in 5G networks.}
      \\ \cline{2-3}
      &\begin{tabular}[c]{@{}c@{}} \cite{cheng2018big} \end{tabular}  & \tabincell{l}{Use DDN to improve the performance of VANETs. }
      \\ \hline

      \multirow{3}{*}{Network security}
      & \cite{sammarco2019unsupervised}  & {Combine artificial intelligence with DDN to monitor malicious behavior. }
      \\ \cline{2-3}
      & \cite{astaras2019deep} & {Propose a data-driven automated security monitoring architecture. }
      \\ \hline

      \multirow{3}{*}{Network optimization}
      & \cite{huang2017data}  & {Introduce the data-driven information plane to provide flexible and intelligent services. }
      \\ \cline{2-3}
      & \cite{chih2017big}  & {Propose a data-driven intelligent radio access network to promote network optimization. }
      \\ \hline

      \multirow{3}{*}{Traffic management}
      & \cite{yao2016novel}  & {For minimizing network traffic by allocating caching resource of content routers.}
      \\ \cline{2-3}
      & \cite{ma2020survey}  & {Evaluate traffic prediction methods based on machine learning in DDN.}
      \\ \hline

     \multirow{1}{*}{Energy efficiency}
      & \cite{zhang2020enabling}  & {Develop a comprehensive solution to improve energy efficiency in IoT.}
      \\ \hline

      \end{tabular}
      \end{table*}


\textbf{1) Network architecture.}
Targeting emerging applications enabled by fifth-generation (5G) networks, Wang {\it et al.} in~\cite{wang2016data} propose a data-based network architecture for collecting users feedback and objective data that can describe a user's subjective experience to enhance personalized QoE in 5G networks.
A DDN framework consisting of the data plane, control plane, information plane, and market plane is considered in~\cite{yin2014big}. The information plane for data acquisition collects topology information, quality of service (QoS) and QoE data, and provides these data to the control plane, which utilizes these data to optimize the network architecture, protocols, resource management and task scheduling.
A mobile network framework enabled by big data analytics is proposed in~\cite{han2017big}, which provides efficient resource allocation, content delivery, and radio access network (RAN) optimization services.
Cui {\it et al.} in~\cite{cui2016big} discuss the benefits of introducing big data to software-defined networks (SDNs), with respect to traffic engineering, cross-layer design, and security assurance.

\textbf{2) Resource allocation.}
In \cite{zheng2016big}, a big data-driven mobile network is discussed, in which users' and operators' data are collected and analyzed to provide best QoE through appropriate resource allocation while minimizing the cost of the infrastructure.
Fang {\it et al.} in~\cite{fang2019data} combine the advantages of information-centric networks (ICNs), SDNs, as well as big data technology to achieve optimal resources allocation, separation of data control and forwarding, and  intelligent content delivery.
To achieve ultra-low latency for computing and caching scenarios, a data-driven model incorporating a data cognitive engine and a resource cognitive engine is proposed in~\cite{chen2018data}.
The data cognitive engine analyzes caching and computing data, and the resource cognitive engine performs resource awareness, such as dynamic storage and computing resources in small cell clouds.
In vehicular networks, Cheng {\it et al.} in~\cite{cheng2018big} show that the vehicle mobility tracking data and the measurement data can be used to evaluate vehicular ad-hoc network (VANET) performance, optimize resource allocation, and design new protocols with big data intelligence.

\textbf{3) Network security.}
Sammarco {\it et al.} combine artificial intelligence methods with DDN and use unsupervised clustering-based procedure to analyze the generated data, thus enabling  malicious behaviors to be monitored in the network~\cite{sammarco2019unsupervised}.
In the field of IoT security, Astaras {\it et al.} introduce a data-driven automated security monitoring architecture~\cite{astaras2019deep}, which is based on reusable security templates, and perform advanced data analysis through deep learning to detect anomalies in each layer of an IoT system.

\textbf{4) Network optimization.}
Huang {\it et al.} in~\cite{huang2017data} introduce the data-driven information plane into the traditional SDN architecture to provide more flexible and intelligent network services.
In~\cite{chih2017big}, Lin {\it et al.} propose a data-driven intelligent RAN by  applying big data in wireless network and machine learning methods to all layers of the communication system to promote intelligent network optimization.

\textbf{5) Traffic management.}
In order to deal with the rapid growth of traffic in the network, Yao {\it et al.}  in~\cite{yao2016novel} suggest that the SDN and content-centric network architectures should be combined to minimize network traffic by solving the problem of caching resource allocation on content caching routers.
In~\cite{ma2020survey}, Ma {\it et al.} focus on the important role of traffic prediction in active network optimization, and evaluate several existing traffic prediction methods based on machine learning in DDN to implement better traffic management.

\textbf{6) Energy efficiency.}
Zhang {\it et al.} utilize the data-driven mechanism to develop a comprehensive solution for large-scale and energy-efficient IoT network~\cite{zhang2020enabling}, which uses reinforcement learning to analyze the data collected by sensor networks in order to improve energy efficiency in IoT.

\subsubsection{Blockchain-enabled Applications and Services in Computer Networks.}

Recently, researchers have investigated and proposed a series of solutions that optimize blockchain technology and apply it to computer networks for different applications and services.
There are several surveys, as summarized in Table~\ref{table:ExistingSurvey}, that comprehensively review these proposed solutions from multiple angles to highlight the benefits of adopting blockchain technology in computer networks.

\begin{table*}[!b]
  \centering
  \caption{Existing works of blockchain-enabled applications and services in communication networks.}
  \label{table:ExistingSurvey}
  \begin{tabular}{m{3.3cm}|m{0.8cm}<{\centering} m{9cm}}
  \hline
  \multicolumn{1}{c|}{Focus}  &  Ref.  &
  \multicolumn{1}{c}{Contributions}
  \\ \hline

  \multirow{3}{3.3cm}{Combined with cloud and edge computing}
  & \cite{uriarte2018blockchain}  & {Compare three solutions based on blockchain in cloud computing, with emphasis of cloud market standards. }
    \\ \cline{2-3}
  & \cite{yang2019integrated}  & {Summarize the current research on the integration of blockchain technologies and  edge computing.}
  \\ \hline

  {Combined with artificial intelligence}
  & \cite{salah2019blockchain}  & { Show a broader perspective of blockchain-enabled AI applications.}
  \\ \hline

  \multirow{14}{3.2cm}{Combined with security management}
  & \cite{ali2018applications} & {Discuss a distributed and untrusted IoT architecture based on blockchain. }
  \\ \cline{2-3}
  & \cite{Zhu2019ACM} & {Summarize the use of DLT in the IoT and the problems that DLT can solve. }
  \\ \cline{2-3}
   & \cite{khan2018iot} & {Investigate the blockchain solution for the security of the IoT layered architecture.}
  \\ \cline{2-3}
  &  \cite{zhu2018identity} & {Focus on the identity management systems based on blockchain in IoT. }
  \\ \cline{2-3}
   & \cite{pohrmen2019blockchain}  & {Investigate heterogeneous networks which take IoT, SDN, fog architecture, and blockchain into one paradigm. }
   \\ \cline{2-3}
  & \cite{yang2019survey}  & { Investigate the application of blockchain in security of multi-layer network.}
  \\ \cline{2-3}
  &  \cite{salman2018security}  & { Investigate some blockchain solutions for several types of security services.}
  \\ \hline

  \multirow{10}{3.3cm}{Applied in IoT networks}
   & \cite{christidis2016blockchains} & {Study on how to use blockchain to promote the sharing of services and resources in IoT. }
  \\ \cline{2-3}
  & \cite{atlam2018blockchain}  & {Describe the benefits and challenges of combining blockchain with the IoT.}
  \\ \cline{2-3}
  & \cite{xie2019survey}  & {Present a survey on some blockchain application in smart cities.}
   \\ \cline{2-3}
  & \cite{al2019blockchain}  & {Focus on the requirements, opportunities, and challenges of introducing  blockchain into different industrial fields.}
    \\ \cline{2-3}
   &  \cite{siano2019survey}  & {Introduce blockchain into energy system to improve the traditional centralized power system.}
  \\ \hline

  \end{tabular}
  \end{table*}


\textbf{1) Combined with cloud and edge computing.} The current states of blockchain for the cloud market, which is dominated by a few providers currently, is investigated in~\cite{uriarte2018blockchain}.
The authors provide some solutions based on blockchain and smart contracts, and then discuss the pending problems for applications in the cloud market.
Yang {\it et al.} in~\cite{yang2019integrated} summarize the current research works of adopting blockchain in edge computing, and survey the enabling techniques in terms of networking, storage, and computing.

\textbf{2) Combined with artificial intelligence (AI).} Several pioneering works have shown the potential benefits of blockchain in AI, such as enhancing data security, making collective decisions, and realizing distributed intelligence.
Salah {\it et al.} in~\cite{salah2019blockchain} present a broader perspective of blockchain-enabled AI applications, including decentralized blockchain-based data storage and management,  decentralized infrastructure, and decentralized AI applications.

\textbf{3) Combined with security management.}
Blockchain is considered in~\cite{ali2018applications,Zhu2019ACM} to build a distributed, trustworthy, and secure architecture for IoT applications .
Khan {\it et al.} in~\cite{khan2018iot} review and categorize common security issues related to the  layered IoT architecture.
Decentralized networks supported by digital ledger technology (DLT), which provides a method to promote trusted interactions between IoT devices, have the potential to achieve security control, identity management, and traceable machine-to-machine transactions.
Zhu {\it et al.} in~\cite{zhu2018identity} focus on identity management systems in IoT, with higher requirements for scalability, mobility, security, and privacy.
They investigate the recent identity solutions based on blockchain, and point out that blockchain can transfer access control and identity management to the edge computing devices close to the identity owners.
Pohrmen {\it et al.} in~\cite{pohrmen2019blockchain} investigate heterogeneous networks incorporating IoT, SDN, fog architecture as well as blockchain technology.
The distributed structure of blockchain can make the entire system more resilient to attacks and single points of failure, thereby reducing the impacts of attacks in SDN-based IoT and cloud-based communications.
Yang {\it et al.} in~\cite{yang2019survey} survey the fundamental issues in the security service architecture, and the usage of blockchain in the data, control, network, and application planes.
Salman {\it et al.} in~\cite{salman2018security} investigate some blockchain solutions for typical security services.

\textbf{4) Application in IoT networks.} Christidis {\it et al.} in~\cite{christidis2016blockchains} study how to adopt blockchain to promote the sharing economy, while using smart contracts to automate multi-step processes in IoT networks.
Focusing on the centralized client/server model in the current IoT, Atlam {\it et al.}  in~\cite{atlam2018blockchain} evaluate the benefits and challenges of combining blockchain with IoT in details.
For the typical IoT scenarios, the advantages of blockchain ({\it e.g.}, distribution, trust-free, transparency, decentralization, and automation) can help improve the services of smart cities, industrial domains, smart grids, {\it etc}.
Xie {\it et al.} in~\cite{xie2019survey} consider the applications of blockchain in smart cities, such as smart citizens, smart healthcare, smart transportation, supply chain management and other issues. 
Al-Jaroodi {\it et al.} in~\cite{al2019blockchain} present the requirements, opportunities, and challenges of introducing blockchain into different industrial fields, where blockchain ledgers can introduce digital identities, security records, smart contracts, and distributed storage to the industrial applications.
Siano {\it et al.} in~\cite{siano2019survey} introduce blockchain into energy systems to improve the traditional centralized power system.
This work reviews the distributed transactive energy systems (TESs) and proposes an interactive energy management architecture for peer-to-peer energy exchange based on distributed ledgers which uses proof of energy (PoE) as a consensus mechanism.

\subsubsection{Existing Works on Blockchain-based Network Architecture}

Many researchers have investigated the integration of blockchain technique into various network architectures. A next generation blockchain network (NGBN) model based on peer-to-peer interactions is proposed in~\cite{Lei_NGBN}. The physical layer and application layer are converged into a single networking layer called the blockchain network layer (BNL), which incorporates encryption, storage, traffic balance, token control, and consensus to ensure secure data transfer with low latency.
In addition, several research works have considered how to combine blockchian with existing frameworks, such as SDN and cloud computing, to optimize the architecture and functionalities of the network in order to provide secure and decentralized services in  typical communication scenarios.

\textbf{1) Blockchain-based SDN architecture.} In order to deal with security challenges in SDN, Weng {\it et al.} in~\cite{jiasi2019secure} propose a secure blockchain-based SDN network architecture, which consists of the data plane, blockchain plane, control plane, and application plane.
The blockchain plane provides functionalities of resource-recording and resource-sharing among multiple controllers in the control plane.
All the application flows and network events associated with the respective network conditions are recorded on the blockchain as transactions.
In the control plane, multiple controllers participating in the underlying blockchain are responsible for recording network data from the application plane and the data plane as transactions into the blockchain.
A kind of Byzantine fault tolerance (BFT) protocols, like the Ripple network~\cite{ripple}, are adopted to guarantee low latency and avoid the temporary forks.

For optical networks, a distributed blockchain-based network architecture is described in~\cite{Yang_BNOPN} along with two distributed multi-controller credible routing (MCR) schemes for software-defined data center optical networks.
Data centers connect to multi-domain elastic optical networks that implement the computing, storage, and optical spectrum resources allocation, respectively.
These domains are software-defined and manipulated by collaborating SDN controllers, and each data center can accommodate trustful cross-domain lightpaths based on the blockchain network.

\textbf{2) Blockchain-based cloud computing architecture.} Sharma {\it et al.} in~\cite{Sharma_BCA} propose a distributed blockchain-based cloud architecture at the edge of the network, which consists of the device layer, fog layer, and cloud layer. The device layer transmits the filtered raw data to the SDN-enabled fog layer, where all SDN controllers are connected in a distributed manner using a blockchain, and each SDN controller is responsible for network management. The fog nodes transmit the processed data to the distributed cloud and device layers, which enables them to access and offload computing tasks to the cloud when computation resources are insufficient. To improve the performance of computation as well as the data transfer and storage in the blockchain, a consensus protocol is proposed to combine the advantages in both Proof of Work (PoW) and Proof of Stake (PoS) consensus mechanisms by 2-hop blockchain technique.

\textbf{3) Blockchain-based IoT architecture.} In ~\cite{Li_ICIOT}, a blockchain-based multi-layer mode is proposed for IoT, which consists of the edge layer and high-level layer. The edge layer is considered as a local area network, where a number of nodes and a leading central node are deployed for providing interfaces to the high-level layer for addressing, allowing bidirectional data transfer, and participating in high-level layer activities. In the high-level layer, all the nodes are data-independent with full replica records exchanged between each other across the blockchain.
They operate with a common interface to the edge layer, independent of the operation of the edge nodes.

\textbf{4) Blockchain-based VANET architecture.} For vehicular communication systems, Zhang {\it et al.} in~\cite{Zhang_BNVANET} present a security blockchain-based architecture of VANET with mobile edge computing.
The architecture consists of the perception layer, edge computing layer, and service layer.
The perception layer, where vehicles and roadside units (RSUs) are connected through blockchain, guarantees high security for the transmitted data.
Computation resources and cloud services are provided by the edge computing layer as well as the service layer. The edge computing layer is responsible for handling a large number of transactions and other computation intensive tasks for the perception layer, while the service layer ensures the security recording of data, including traffic violation history, traffic accident information, {\it etc}.

\subsection{Motivations for Blockchain-empowered Data-driven Networks}

\subsubsection{Challenges of Future Data-driven Networks}
\label{subsubsec:challenges}
\begin{figure*}[b]
\centering
\includegraphics[width=0.9\textwidth]{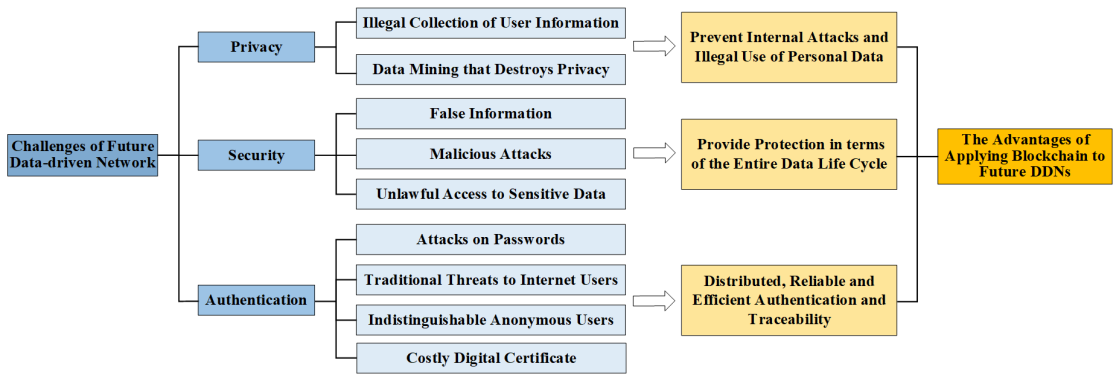}
\caption{Challenges of future data-driven networks.}
\label{fig:challengesDDN}
\vspace{-0.2in}
\end{figure*}

As the types of data and services keep growing, the concepts of data-driven services will continue to evolve, which results in increasing scale and performance requirements of future DDNs, as well as other challenges, as illustrated in Fig.~\ref{fig:challengesDDN}.

\textbf{1) Privacy:} For data-driven services, the analytics of user's data is conducive to support personalized applications.
With the increasing variety of data collected and stored in the future DDNs by Internet service providers and data-mining companies, how to protect users' privacy while providing intelligent services becomes a great challenge.
According to the research in
\cite{III4, III8},
there are many issues that affect users' privacy. Massive user data, which are collected, transmitted, and analyzed in the network, may reveal informative results and carry immense values.
However, users may not be informed about what their data are used for, nor how to protect their privacy or secure their personal information.
There have been several disclosed cases in which illegal operations have been performed without authorization or users' knowledge, resulting in severe personal information leakage and punishable offence to public safety.
On the other hand, most data mining uses linking, reconstruction, and inference operations, which may potentially breach users' privacy.
Conventional privacy protection approaches ({\it e.g.}, encryption, anonymity, client personalization, {\it etc.}) focus on enterprise networks by addressing the vulnerabilities at network gateways and access portals~\cite{Kobsa2007Privacy, Hua2019CINEMA}. These operations are performed by the centralized storage, computing, and data processing.
However, these methods cannot prevent the leakage of user privacy caused by internal malicious attacks.
With mobile users using online applications that generate/consume data in a decentralized manner, previous privacy protection mechanisms may no longer be adequate for personal data and privacy protection.
Future DDNs should solve this problem in a complete, consolidated and secure paradigm instead of the current patchwork and isolated solutions deployed by individual Internet service providers.

\textbf{2) Security:}
Security has been a matter of great concern in computer networks due to the proliferation of many forms of malicious attacks \cite{III11, III15}, {\it e.g.,} de-anonymization attacks, sniffer attacks, distributed denial-of-service attacks (DDoS), and Domain Name System (DNS) attacks.
In future DDNs, the variety and complexity of data resources and formats, as well as their respective transmission, storage and analytics requirements, elevate the security issues to a new height with new challenges.
Data-driven applications are designed on the premise of effectively utilizing various data to provide intelligent and customized services. False information and malicious attacks can destroy the credibility of the data, resulting to unacceptable QoS or even hamper the normal operating of the devices and terminals. Furthermore, unlawful access to sensitive and private data by hostile hackers may bring unimaginable losses for the data owners and the public.
Existing security solutions for computer networks are designed to counteract specific malicious attacks in a certain network environment~\cite{Xu2019Efficient, He2020DNS}. The current approach of reacting to the ever-emerging novel malicious attacks with tailored solutions leads to delays in the introduction of effective actions and tedious patching on the existing networks.
New security paradigms need to be devised for future DDNs to provide powerful protection in terms of the entire life cycle of the data.

\textbf{3) Authentication:} For DDNs with multiple service providers, infrastructures, and users, authentication is very important for mutual identification and data confidentiality.
However, there are still several problems in authentications
\cite{III16, III20}.
Currently, the most widely used form of user authentication in modern computer systems is password, which is vulnerable to password leakage as well as the password-cracking and social-engineering attacks.
In addition, the traditional threats to Internet users include message replay, dependence on trusted third parties which might be compromised, and man-in-the-middle (MITM) attacks.
All these threats can prevent legitimate users from being successfully authenticated or allow illegitimate users to be authenticated.
As  to  the  existing  solutions,  in  some  network  scenarios,  anonymous authentication may provide secure authentication and access control~\cite{Ma2010Privacy}.
However, anonymity also brings some problems.
Many studies have focused on the implementation of pseudonymization technology (PT) by introducing trusted authorities~\cite{Lu2012Pseudonym}, which are easily affected by the single point of failure.
In addition, considering that anonymous users are often indistinguishable, it is difficult for the network service provider to punish misbehaving anonymous users.
When individual authentication is required, digital certificates are often used for this purpose~\cite{Wohlmacher2000Digital}, which may consume a significant amount of network overhead for computation and certificate validation.
Therefore, the efficient implementation of effective authentication methods becomes very important.

\subsubsection{Advantages of Applying Blockchain to Future Data-driven Networks}
\label{subsubsec:applying blockchain}

DDNs require massive storage and processing of data in the networks, for which traditional solutions such as database technologies are applicable. Compared with representative database technologies such as Oracle and MySQL, blockchain has obvious merits. As a distributed ledger, blockchain can effectively avoid data leakage and single point of failure caused by authorized administrators, while its anonymity can protect the privacy of data owners to a certain extent.
The blockchain only retains the two database operations of data reading and adding, and the process of adding data must be verified by the consensus protocol, which ensures the transparency and integrity of the entire database.
Therefore, applying blockchain to the future DDNs can uniformly solve the above challenges in terms of privacy, security and authentication.
Specifically, blockchain technology provides a promising solution that can address many of the challenges in future DDNs while bringing several potential benefits as follows:

\textbf{1) Security and privacy:} Existing works have discussed about how to apply blockchain to improve security and privacy in the networks from different points of view

\cite{III23, III24}.
First, blockchain nodes are decentralized and supportive to network robustness.
Even if some nodes in the network are compromised by various attacks, other nodes can work normally and the data will not be lost.
This feature of blockchain increases the overall network robustness compared to existing centralized and distributed data systems.
Second, the transparency of blockchain can be utilized to make the data flow in the network completely open to users. This provides traceability to users' personal data usage, thus informing the users how their data is used.
Third, the immutability of data in the blockchain increases the reliability of activities in the network and enhance the mutual trust between users and service providers.
Finally, pseudonym of blockchain helps network users to keep their real-world identity hidden for privacy protection.

\textbf{2) Data and model sharing:}
In order to support massive data and application requests for sharing data and information in a secure and effective manner, applying blockchain in future DDNs can protect network data from tampering, and effectively preserve the privacy for network users.
Moreover, blockchain is more transparent and secure in data sharing.
By better protecting the privacy for users' data and the security of users' application, BDNs can provide better personalized services with more add-on values.

\textbf{3) Credibility and malicious operation tracing:}
With the immutable and distributed advantages of blockchain, future DDNs can increase the credibility of the network. Malicious operations and mendacious messages can be traced back by all the participants that can access the transaction records saved in the blockchain. Cloud-based services can store the credit value information of the service providers, which enables both the users and service providers to correctly verify the legitimacy of the operations.

\textbf{4) Enhanced decentralized solutions:} As a distributed ledger of transactions, blockchain technology is quite suitable for peer-to-peer interaction and decentralized intelligent services, such as federated learning-based applications.
As long as the nodes of the system are running compatible consensus mechanism or protocol, they can access the transaction records while not being able to change any of them without being noticed. Thus, it can provide a more secure and convenient method compared with current decentralized solutions for data sharing and cooperation among independent nodes in a large scale network deployment.

\section{An Overview of Blockchain and Data-driven Networks}
\label{sec:overview}
\subsection{Blockchain}

A blockchain employs decentralized digital ledgers that are maintained by peer nodes. Generally speaking, the architecture of blockchain involves six layers, {\it i.e.}, the data layer, network layer, consensus layer, incentive layer, contract layer, and application layer,  as illustrated in Fig.~\ref{fig:layers}~\cite{yuan2018blockchain}.
Each layer has specific core functions and key technologies as described below. As for more background of blochchain including the characteristics and operation types, please refer to the online supplemental file.

\begin{itemize}
\item {\it Data layer}:
The data layer is the lowest layer in the blockchain architecture. Its key technologies include Merkle tree, asymmetric encryption, timestamp, digital signature, and hash function.
The data layer constructs the basic data structure of a blockchain for organizing and storing data~\cite{wu2019comprehensive}.
A typical blockchain data structure consists of a header part and a body part.
The header part contains meta-information such as version information, hash value of the previous block header, timestamp, nonce, Merkle root of the contained transactions, and target difficulty used to calculate the next block.
The body part stores a Merkle tree of the verified hash data, which enables the blockchain to verify the existence and integrity of data efficiently and securely.
\begin{figure*}[b]
\centering
\includegraphics[width=0.6\textwidth]{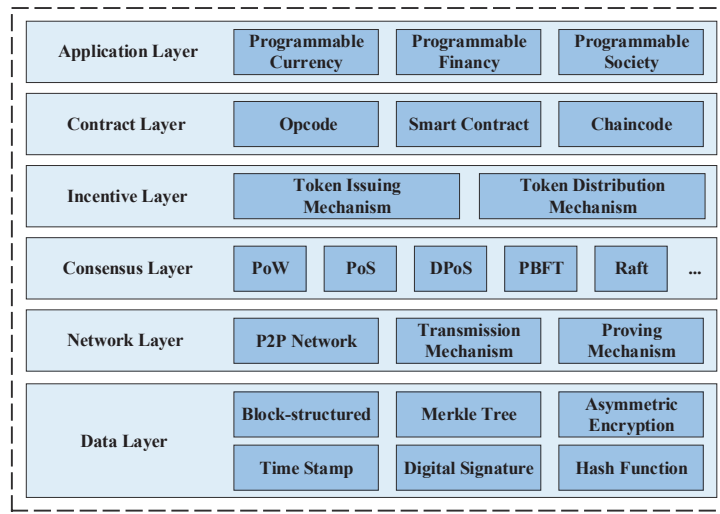}
\caption{The six-layer blockchain protocol stack.}
\label{fig:layers}
\vspace{-0.2in}
\end{figure*}

\item {\it Network layer}:
Blockchains operate in a peer-to-peer paradigm.
The main task of the network layer is to support the exchange of information between peer nodes in the network, while ensuring the security and privacy of the data. In a peer-to-peer network, the node that generates a transaction broadcasts the transaction to its neighboring nodes.
Each node that receives the transaction then verifies it according to the corresponding check list.
Only authenticated transactions are forwarded by the node.

\item {\it Consensus layer}:
The consensus layer encapsulates the consensus algorithm for the distributed nodes.
It guarantees the data consistency and fault-tolerance of the shared ledger.
There are some well-known consensus algorithms such as PoW, PoS, Delegated PoS (DPoS), Practical Byzantine Fault Tolerance (PBFT), Raft, {\it etc}.
PoW requires nodes to compete for opportunities to append blocks on the ledger through mathematically difficult calculations~\cite{wu2019comprehensive}.
It relies on the computing power of nodes.
In a PoS system, the creator of the next block is determined by the number of assets a peer holds and the duration since the peer created a block last time.
Compared to PoW, PoS requires far less resources to run, but decreases the liquidity of cryptocurrency.
DPoS is a variant of PoS, which generates block producers  in a round-robin order.
It sacrifices the complete decentralization characteristic to provide high throughput and scalability.
PBFT uses the democratic mechanism of majority-rule to select leaders and keep accounts.
It is a fault-resistant, fast, and long-lived mechanism.
However, a large number of nodes will consume a heavy communication overhead.
Raft is an easy-to-understand general consensus protocol that uses the ``leader and follower'' model.
As a pluggable consensus module, Raft is currently supported by some popular blockchain architectures such as Hyperledger Fabric.

\item {\it Incentive layer}:
The incentive layer, which integrates economic rewards into a blockchain system, mainly appears in public blockchains.
In such a decentralized system, nodes are actually self-interested and their fundamental purpose of participating in data validation and book-keeping is to maximize their own benefits.
Using tokens to reward nodes that follow the rules is the most extensive incentive mechanism in the blockchain system.
The incentive layer formulates the token issuing and distribution mechanism, which determines the total amount and circulation of tokens.

\item {\it Contract layer}:
The contract layer encapsulates various Opcodes, Chaincodes, and smart contracts.
In blockchain systems, smart contracts are predefined commitments and rules~\cite{watanabe2015blockchain}, which are executed automatically when some predetermined conditions are met.
Smart contracts enable trusted transactions without a third party, and these transactions are traceable and irreversible.
Opcodes and Chaincodes specify the details of trading and processes, thereby increasing the autonomy and programmability of the blockchain network.

\item {\it Application layer}:
As a basic technology, blockchain has been applied in multiple fields, such as financial market, healthcare systems, smart city, and energy market. The highest layer of blockchain includes various applications to provide secure, distributed, and customized services.
In the long run, blockchains will bring more solutions to future networks when combined with advanced data and communication technologies.
\end{itemize}

Researchers have continued to contribute to the technologies and protocols associated with blockchain \cite{christidis2016blockchains}.
A comprehensive survey on the security and privacy aspects of blockchain is presented in~\cite{Zhang2019ACM}.
Neudecker {\it et al.} in~\cite{neudecker2018network} associate the attacks and security requirements with network layer design, and investigate the requirements and adversary model of the network layer design.
The impacts of consensus and incentive mechanisms on the consensus participants are investigated in~\cite{wang2019survey} from a game-theoretic perspective, which highlights how the consensus mechanisms affect the emerging applications of blockchain networks.
A vademecum has been made in~\cite{belotti2019vademecum} to guide designers.
The authors not only comprehensively introduce the existing blockchain platforms, but also propose the key requirements when evolving from the permissionless blockchains to the permissioned blockchains.

\subsection{Future Data-driven Networks}
\label{subsec:future_ddn}

The rapid development and widespread of data-driven applications have brought promising opportunities to reshape the Internet architectures as well as operation and optimization solutions, thanks to emerging data analytic capabilities that can uncover the knowledge and statistical patterns hidden in massive data~\cite{yin2014big}.
Data-driven functionalities may become some of the most important features for future computer networks, particularly to efficiently handle the skyrocketing user traffic while leveraging the huge amount of data generated and digested inside the networks to improve the effectiveness of network management, resource allocation, and security control.

\begin{figure*}[!b]
\centering
\includegraphics[width=0.5\textwidth]{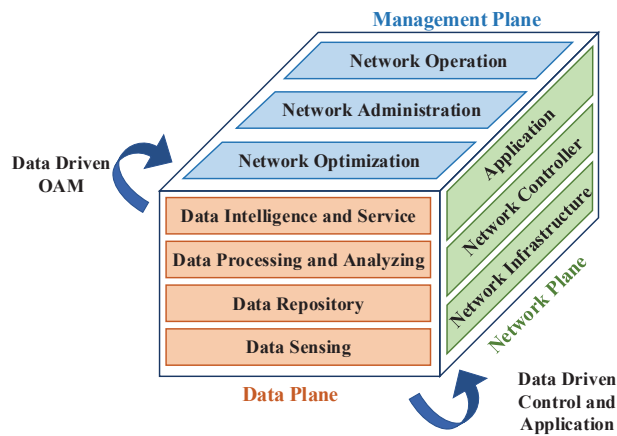}
\caption{The layers in future DDNs.}
\label{fig:data-driven}
\vspace{-0.2in}
\end{figure*}

\textbf{1) Layered architecture in future DDNs:} The architecture of future DDNs consists of three planes, {\it i.e.}, the data, management, and network planes, as shown in Fig.~\ref{fig:data-driven}.
Unlike the traditional vertical layered model, the proposed 3-dimensional (3D) model also describes the interactions within and between the three planes.

The data plane consists with four layers, including the data sensing layer, data repository layer, data processing and analyzing layer, and data intelligence and service layer.
The data sensing layer collects various data, including user traffic, network  performance indexes, as well as the network management signalling and operational directives, from the network entities.
Such real-time information is critical to enable the use of advanced data analytic methods to reveal the true network status and problems.
The data repository layer stores all the collected data. The cloud architecture could be used for such purposes.
Then, the data processing and analyzing layer processes the data and outputs the necessary information to enable the data intelligence and service layer to solve network problems and support optimal decision-making.
This process enables effective resource allocation strategy for network self-optimization, as well as end-to-end network intelligence as discussed in the context of the other two planes.

The network plane consists of the network infrastructure, controller, and application layers.
The network infrastructure layer includes all kinds of network entities in various typical communication scenarios.
The network controller layer works with function modules in distributed network nodes and centralized management servers and is responsible for policy making and dispatching.
The application layer provides various network applications and services for the users, such as network reconstruct, secure transmission, and QoE.

The management plane has three layers that are responsible for network operation, network administration, and network optimization, respectively.
Taking advantage of data analytics and end-to-end network intelligence in the data plane, a series of automatic operations including smart maintenance, troubleshooting, configuration, and optimization functionalities can be implemented in the management plane.
It can further provide elaborate fine-grained network control according to practical requirements and real-time feedback of network status to improve QoE for users.

\textbf{2) Characteristics of Future DDNs:} With three planes orchestrating effectively, future DDNs are expected to have following advantages in support of intelligent services with desirable QoE.

\begin{itemize}
\item {\textit{Intelligent and autonomous control}: Future DDNs with highly sophisticated data computing and analytic capabilities can support autonomous network operation and maintenance.
With the help of virtualization and ever-increasing computing capacity, network elements become more intelligent and thus can react quickly to changes in the network environment.
Thus, future DDNs can update the network parameters autonomously in real time during network operation.}

\item{\textit{Low cost and high efficiency}:
Data from all network entities could be stored and analyzed efficiently to optimize network operation, which can  decrease the operating expense (OPEX) and capital expenditure (CAPEX).
Intelligent self-maintaining scripts and algorithms could be designed for joint resource allocation and smart network operation to free engineers from cumbersome manual tasks.
Furthermore, due to the growing computing capacity and hence intelligence  of network elements, many decisions can be made and executed in a distributed manner, which can potentially greatly decrease the OPEX.}

\item{\textit{Low latency and real-time operation}: In future DDNs, the data plane can perform the time-consuming data processing and model training offline, thus providing the tuned model to the management plane for quick reaction to the changes of the the network or user requirements. The distributed and cooperative operation of the network entities in the network plane may also help to realize low latency for the end users. }

\item{\textit{Heterogeneity and scalability}:
Future DDNs are flexible, extensible, and scalable for the ever-increasing network coverage and heterogeneous types of accessing devices.
Moreover, network nodes and servers, which may be implemented across platforms, can collaborate effectively in data collecting, processing, and analyzing by highly intelligent information exchange on the premise of powerful security, privacy and authentication protections. }
\end{itemize}

\section{Framework of Blockchain-empowered Data-driven Network}
\label{sec:frmwk_of_blkchn_ddn}

While future DDNs promise to provide efficient, flexible and intelligent services for an ever-increasing number of users and their terminal devices, there are many challenges that may hamper the fast evolution of present-day networks towards DDNs, as we have addressed in Section~\ref{subsubsec:challenges}. By leveraging blockchain technology, we believe that promising BDN solutions can be found to address these challenges, as indicated by the discussions in Section~\ref{subsubsec:applying blockchain}. To facilitate research towards these solutions, we present in this section a general framework of BDNs, which provides a scalable and universal architecture for integrating blockchain into future DDNs.

\begin{figure*}[!b]
\centering
\includegraphics[width=1\textwidth]{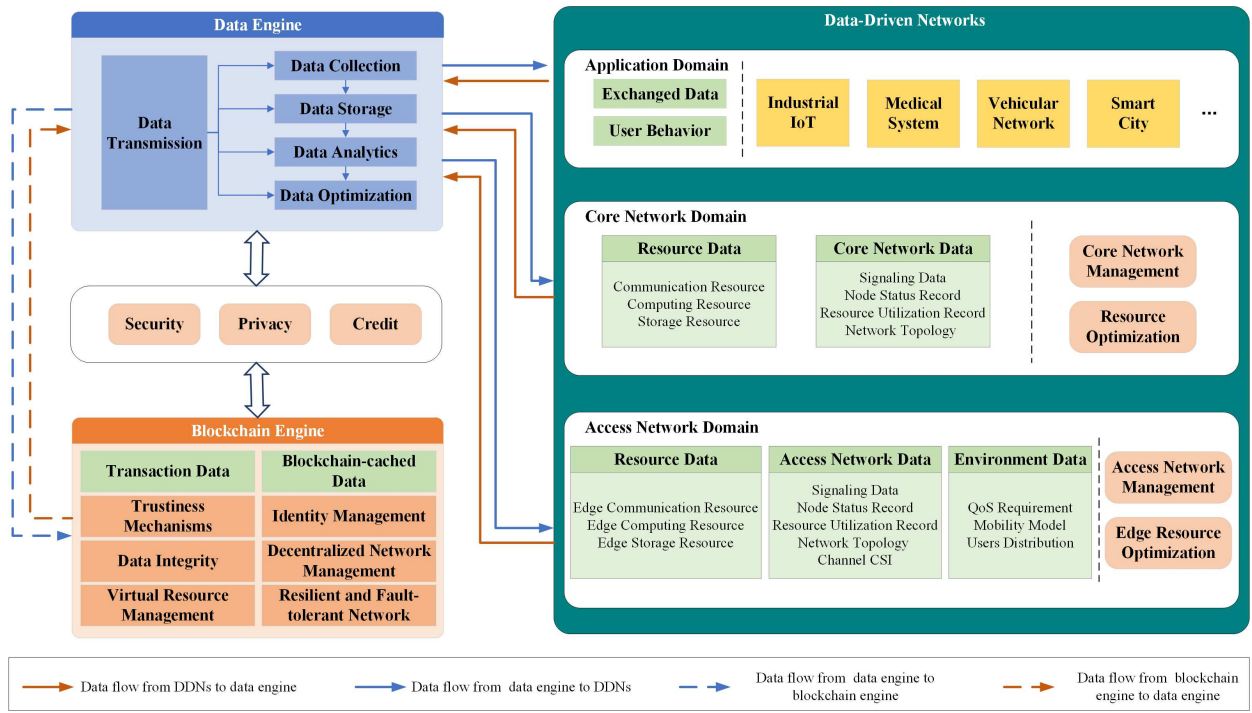}
\caption{Architecture of blockchain-empowered data-driven networks.}
\label{fig:architecture}
\vspace{-0.2in}
\end{figure*}

As shown in Fig.~\ref{fig:architecture}, our framework of BDNs is based on the future DDN described in Section~\ref{subsec:future_ddn}. While the data engine collects as well as processes massive and various data in different domains of the network, the blockchain engine interacts with the data engine by providing secure data storage, private data sharing, and decentralized network operation.

\textbf{1) Network domains:} We divide the network into three domains, namely the access network domain, core network domain, and application domain.

\textit{The access network domain} is where a myriad of terminals connect to the network by different communication techniques through wired or wireless links. There are also various edge network nodes, such as base stations and accessing points, edge switches and routers, gateways and edge data centers, cooperating together to perform access control, resource allocation, and data processing, with huge amount of generated and digested data including resource utilization status, operational node performance, and environment monitoring results.

\textit{The core network domain} consists of large number of powerful routers and high-speed optical links, distributed cloud computing centers and data storage repositories, and facilitates the main functions of network management, resource optimization, and the signalling and data packets.

\textit{The application domain} provides interfaces for various kinds of services and user applications,
such as industrial IoT, medical system, vehicular network, and smart city.
The application layer interacts with the core network and access network layers via application-control interfaces of the nodes to support different services flexibly and intelligently.

\textbf{2) Data engine and blockchain engine:} The core modules of the proposed BDN framework are the data engine and blockchain engine. All the network domains interact with the data engine, which is responsible for various operations related to all kinds of data. And then, the data engine and blockchain engine exchange data and parameters with each other to coordinate operation and optimization of the network.

\textit{The data engine} is composed by several core components. These modules are physically distributed in various nodes and devices across the network.
The data collection module, which is equipped in all the users' terminals and network nodes, collects various data including user behavior, resource utilization status, traffic data, device performance, management signalling and operational messages from the network in a timely manner.
It also monitors and senses external network environment, such as geographic location and electromagnetic interference.
The collected data is stored in the data storage modules of edge and core data centers based on the cloud architecture.
The data analytics module, equipped with strong computing ability as well as a comprehensive library of data-mining and data analytic algorithms, is called upon to digest these stored data in different dimensions according to the requirements from the data optimization module. All the nodes with available computing and storage capabilities, including powerful servers in data centers, routers, edge nodes, and user terminals, can be equipped with a data analytics module, enabling them to execute data analytics tasks independently or cooperatively. There are interfaces for sharing information among these modules and various nodes, which is an important standardization work for the BDNs. The close cooperation among these modules in the data engine guarantees integrity and efficiency of data-driven services and operations.
TThe Merkle tree and block hash are used to secure verification of data in a large dataset, and help to verify the consistency of the data. A set of nodes store all the blockchain data, which can be easily synchronized and maintained in the event of an individual node failure.

\textit{The blockchain engine} takes advantage of the decentralized, secure, and private features of blockchain, and is supported by each node inside the blockchain network. It enables a series of network functions, including trust mechanisms, data integrity, virtual resource management, attack-resilience and fault-tolerance, which will be discussed in Section~\ref{sec:app_survey} and  ~\ref{sec:manage_survey} with more details.
It can further support secure and private data sharing, storage, and processing as well as decentralized network operation according to requirements and feedback, which can provide users with better QoE.
In addition, the blockchain engine interacts with the data engine in a secure, effective, and trustful manner.
The blockchain engines in the network nodes collaboratively select a node according to the applicable consensus algorithm and authorize it to create a new block to encapsulate the transaction data with a specific timestamp generated over the network. In this way, a myriad of data can be shared among completely decentralized nodes and entities.

With the data and blockchain engines, the proposed framework of BDNs can form the basis for the design and implementation of future BDNs to enable intelligent and secure distributed network operations.
Considering the explosive increase of network data and their myriad of values, BDNs are capable of performing comprehensive data fusing and information extraction over the  huge amount of collected raw data. By correlating various influencing factors and network performance, the data engine can deduce the causality and logic behind these data with the help of advanced data analytics and machine learning techniques by taking advantage of the ever-increasing computing capacity of the nodes.
Then, it can optimize network operation intelligently and efficiently, especially when the network behaviors are complicated and a massive number of network parameters need to be managed. Furthermore, with proper authorization, the output of the data engine can be made available to third-party application providers through standardized interfaces, thus supporting diverse and high-quality services to users.
The blockchain provides decentralized control that is particularly fitting for the condition that organizations or individuals are situated in different geolocations. The individual nodes in future BDNs can delegate or vote their representative authority for decentralized control rather than limiting it to a few authorized nodes. Data integrity can also be guaranteed by a blockchain, which introduces a set of protocols to verify the participating nodes as well as new transactions. After the majority of nodes have reached a verification consensus, the new transactions are recorded and saved in the new block within the distributed architecture to improve the data security and integrity. The blockchain engine operates harmoniously and effectively with the data engine to protect the secure data storage and trustful network management.

Under the proposed framework of BDNs, data from all the network entities at various levels could be stored and processed to optimize network management and operation, which can reduce both OPEX and CAPEX. Blockchain with digital signature and hash functions are applied to guarantee the data integrity and immutability in the network. In this way, the future BDNs can provide efficient solutions to achieve optimal network control and data-driven applications, especially when the network environments and scenarios are complicated and frequently changing with various users requirements.

\section{Application of Blockchain-empowered Data-driven Networks}
\label{sec:app_survey}
\subsection{Decentralized Data Sharing and Data Privacy}

In future networks, massive data generated by connected devices and network equipment provide promising possibilities for improving the QoS of the emerging data-driven applications through data sharing. However, these data are scattered in different systems, and controlled by the respective service providers.
How to enable secure data sharing and analyzing in a decentralized manner is an important issue that may be addressed through the application of BDNs.
Meanwhile, privacy is one of the key issues in data sharing. Conventional privacy technologies, such as encryption, authentication, and role-based access control, may not be sufficient to satisfy the efficiency and security requirements in future computer networks.
As BDNs may provide effective solutions to privacy issues, many researchers have conducted pioneer works from different directions.

A data sharing system based on blockchain can greatly simplify the data acquisition process, control access to the stored data, track the use of data, and facilitate users' data ownership and privileges.
A blockchain-empowered data sharing architecture is designed in ~\cite{Lu_BNFL} for distributed multiple parties.
These parties agree to share their own data so as to implement a collaborative task together.
Considering the common distributed data sharing framework involving multiple parties, all the IoT end devices are interconnected within a permissioned blockchain, which is maintained by the super nodes implemented with computing and storage resources.
Using the federated learning method, the permissioned blockchain stores the federated data model learned over decentralized parties instead of all the raw data, which can trace the usage of data for further auditing while ensuring data security and privacy.
How to improve the mapping of raw data to federated data under the limited resource constraint of IoT devices is a key problem in data sharing.

In order to maximize the data collection ratio as well as geographic fairness,  a joint deep reinforcement learning (DRL) and blockchain-based secure data sharing framework is proposed by Liu {\it et al.} in~\cite{Liu_BNDS_DRL}.
A fully distributed DRL scheme is designed to help each mobile device to sense nearby environment to achieve maximum data collection amount, geographic fairness, and minimum energy consumption.
Ethereum blockchain is utilized to build a trusted platform for mobile devices to share data without any third-party organization.
Mobile devices in the blockchain collect data through a multiagent DRL-based method, and transmit the encrypted data with respective private key and digital signature to ensure data safety.
Further improvement can be carried out from the aspect of improving the node intelligence using DRL.

Chen {\it et al.} in~\cite{Chen_DS_BN} propose a blockchain-based data sharing incentive mechanism for the Internet of vehicles, and design a data quality-driven auction to perform the negotiation among data buyers and sellers for guaranteeing high-quality data and maximizing social welfare.
A smart contract is designed for traffic data sharing among vehicles.
A tamper-resistant consortium blockchain is introduced to securely store the on-chain data such as timestamp and shared information after completing the data sharing ensuring the security and scalability of the algorithm.

Many researchers utilize blockchain to realize the secure management and storage of distributed data with encryption to facilitate sharing.
A data storage and sharing framework is presented in ~\cite{IVC23}, which consists of the InterPlanetary File System (IPFS), the Ethereum blockchain, and the attribute-based encryption (ABE) technology.
Along with controlling the data that they own, users also encrypt the shared data through specified access control schemes.
Moreover, encrypted keyword indexes are established for shared files, and the smart contract ensures that users need to pay a service fee only when the cloud server returns the correct result.
However, it is challenging to reduce the costs associated with users' attribute revocation and access policy updating.
For secure blockchain-based online storage, Fukumitsu {\it et al.} propose a scheme without any third-party central server~\cite{IVC21}.
Data are secretly divided into several parts and sent to the nodes through secret sharing, which makes it difficult for attackers to obtain all the user data. The distributed storage nodes remove the need for a centralized repository.
A data sharing mechanism called {\it Meta-Key} is proposed in~\cite{IVC22}, which shares encrypted data based on the distributed storage compatible architecture of blockchain.
It allows users to encrypt data with their own public keys, and store the keys to the dedicated storage nodes in a blockchain network.
The system can be improved by adding erasure codes to further enhance the security and reliability of the data cipher-text.
In~\cite{IVC24}, Raman {\it et al.} utilize a new combination of private key encryption, distributed storage, and Shamirs secret sharing schemes to distribute transaction data, with data integrity ensured by specific encoding schemes.
In addition, the Shamirs secret sharing scheme is used for hash value and dynamic region allocation to ensure secure data sharing under the premise of data integrity.

In healthcare, the authors in~\cite{IVC25} present a conceptual design employing cloud-assisted blockchain technology for users to share their personal health data securely.
They focus on continuous dynamic data, which account for most of the data generated by wearable devices and mobile devices, and integrate blockchain and cloud storage technologies to collect and share the dynamic personal health data.
Jin {\it et al.} investigate a novel secure and privacy-preserving medical data sharing mechanism in~\cite{Jin_BNMDS}, which incorporate two types of blockchain techniques, namely, the permissioned and permissionless blockchains.
In~\cite{jiang2018BlocHIE}, different sharing requirements of medical data from medical institutions and individuals are analyzed to drive the design of a healthcare information exchange platform composed of two loosely-coupled blockchains.
The transaction packing algorithm in the two blockchains can effectively increase the system throughput and the fairness of data sharing.
{\it MeDShare} is proposed as a blockchain-based medical data sharing system in~\cite{Xia_MED}.
A model based on blockchain for data sharing between cloud service providers  is designed  to ensure immutability and non-tampering of data.
Also, smart contracts and access control schemes are introduced to control and track the process of data sharing for detecting violation of permissions on data.
This mechanism can safely achieve data provenance and auditing while ensuring user privacy.

A decentralized personal data management platform is proposed in~\cite{zyskind2015decentralizing} to protect data privacy by utilizing the blockchain technology, targeting fine-grained access control management with out-of-chain storage.
Two types of transactions, namely transactions of access control and transactions of data storage and retrieval, are recorded by blockchain.
Individual users can control their data, while service organizations can access the data to provide personalized services after legal authorization.
One of the main contributions of this platform is to use protocols and encryption to overcome the public nature of blockchain; however, it does not mitigate threats coming from malicious nodes in the network, which can increase their reputation and then carry out attacks.

Focusing on privacy requirements in IoT networks, an end-to-end data privacy-preserving framework named {\it PrivBlockchain} is described in ~\cite{loukil2018towards}, which utilizes smart contracts to enforce privacy requirements and regulations between data owners and data consumers.
By deploying core components of {\it PrivBlockchain} such as smart contracts, gateway nodes, {\it etc.}, with appropriate transaction protocols, IoT resources can be added, stored, and shared without violating the end users' privacy, thereby protecting privacy throughout the entire IoT data lifecycle.

The authors in~\cite{kaaniche2017blockchain} propose a hierarchical blockchain-based data usage auditing architecture that relies on auditable contracts in the blockchain to provide controllable and transparent data retrieving, sharing, and processing.
This architecture consists of three entities: a data owner, a data controller, and a data processor.
The authors make assumptions about the actions of both data owners and service providers, and design an auditable contract and encryption mechanism that not only protects user privacy, but also forbid unauthorized entities to process the data.

Chen {\it et al.} in~\cite{chen2019deplest} propose a distributed partial ledger storage technique based on blockchain to protect user privacy in social networks.
This technique combines blockchain technology with database file storage method, and thus can store sensitive information securely.
In order to save the computing power of user equipment, the proposed scheme does not need to synchronize the entire ledger, and use proof of communications as a consensus mechanism.
However, it does not perform well when the network is small ({\it e.g.},~less than 1000 users) or has limited computational resources.

In order to protect multimedia data privacy and provenance, Vishwa {\it et al.} design a decentralized data management platform to store, query, share, and audit multimedia data in~\cite{vishwa2018blockchain}.
The proposed platform combines blockchain encryption with a cloud storage solution called data-lake to ensure that multimedia owners have knowledge of the collection and utilization of their data without relying on a third party.
The authors make full use of the trust-free nature of blockchain, and use smart contracts to program the authenticity checks and authorization rules into the system to ensure user rights; however, overall characteristics of the system have not yet been fully verified.

Using searchable symmetric encryption, it is now possible to implement secure encrypted data search in a distributed database.
Jiang {\it et al.} in~\cite{jiang2019Privacy} propose a bloom filter-enabled search protocol to address data privacy issues in multi-keyword search.
It can not only provide privacy-protected multi-keyword search in the blockchain but also improve the efficiency of data search.

Privacy protection in e-health systems has attracted wide attention with several research works.
Al Omar {\it et al.} in~\cite{al2019privacy} design a patient-centric medical data management system, which uses blockchain as the storage technology.
The system adopts a protocol called {\it MediBchain} to ensure data privacy, accountability, authenticity, and integrity.
The patients and doctors act as data senders, and their data are first encrypted in the {\it MediBchain}.
The private accessible unit (PAU) is an intermediate unit between blockchain and users, which is used to authenticate the identity of the data senders and recipients.
User data is stored in the blockchain.
Each transaction in the blockchain returns a transaction identifier that helps the user to access the data in the future.
The authors in~\cite{dagher2018ancile, magyar2017blockchain, nortey2019privacy} also focus on the privacy of electronic health record (EHR).
Specifically, Dagher {\it et al.} in~\cite{dagher2018ancile} propose a framework named {\it Ancile}, which is built on the Ethereum platform.
Smart contracts and encryption technologies are used in this framework to enable data security, access control, privacy and interoperability for EHRs.
Magyar {\it et al.} in~\cite{magyar2017blockchain} consider how blockchain helps to solve secure data storage problems while providing patient data in a standardized manner.
The distributed EHR management framework in~\cite{nortey2019privacy} protects user privacy during data collection, management, and distribution, by using blockchain for managing access control and EHR storage.

\subsection{Secure Data Storage}

In future BDNs, various data will be stored inside the networks with diverse security requirements.
As an immutable transaction ledger, blockchain can provide secure and distributed storage while enforcing data integrity via proof-of-retrievability schemes.
Data can be stored not only in the blockchain, but also off-chain, depending on the importance of the data and practical requirements.

{\it Storj} is an open-source implementation for blockchain-based cloud storage to provide data security and integrity in distributed applications ~\cite{wilkinson2014storj}.
The data is broken up and stored in peer nodes across the network, while blockchain stores metadata information about where to find the data pieces.
{\it Storj} returns control of cloud data to users, and improves security and privacy by running untrusted and fault-tolerant systems on trusted data providers.
When a user needs to access the data, the blockchain is queried and then returns the required metadata to retrieve the original data.
However, it is not feasible to use the Bitcoin blockchain to store metadata currently, so the system uses a technology named Florincoin instead.
A system named {\it BlockDS} is designed in~\cite{do2017blockchain} to provide secure data storage and keyword search, which consists of three parts: distributed data storage, anonymous access control, and private keyword search.
Similar to {\it Storj}, data references, rather than encrypted data themselves, are stored in the permissioned blockchains, in which different clients have different data access authorities.
The keyword searching component is modeled as a smart contract in the blockchain, and data customers who have been authorized by the anonymous access control layer are allowed to search in the cloud storage without downloading the whole dataset.
Therefore, the system can outsource the data storage to service providers to achieve a smooth distributed network.

Considering that users often have a large amount of under-utilized storage resources in their devices, a secure decentralized storage framework named {\it BlockStore} is proposed in ~\cite{ruj2018blockstore}.
Users can rent out their available storage resources, and the system collects and distributes them to the tenants.
The difference between {\it BlockStore} and traditional solutions is that it provides powerful auditing capabilities through smart contracts to ensure data security and prohibit double rentals.
This solution is just a basic framework for blockchain storage and can be expanded in multiple directions, such as adding incentive mechanism or designing a more efficient wallet data structure.

In order to improve the security of cloud storage and reduce transmission delay, a blockchain-based security architecture is presented in~\cite{li2018block} for distributed peer-to-peer cloud storage.
The uploaded files are divided into encrypted data packages and then transmitted to the nodes in the peer-to-peer network randomly.
A genetic algorithm is integrated into the architecture to solve the problem of data packages placement between multiple users and data centers in a distributed environment.
As a trading mechanism between storage service consumers and providers, blockchain stores file locations such as file hash and file URL, and uses Merkle trees to ensure data integrity.
Through performance and security analysis in a multi-user network with multiple data centers and blockchain, this architecture has shown to achieve a lower file loss rate and transmission delay.

{\it Mchain} proposed in~\cite{zhao2018mchain} has a two-layer structure to improve the security and latency of the virtual machine (VM) for data storage in the infrastructure-as-a-service (IaaS) cloud architecture.
The authors use the same trust assumptions as the original blockchain network, that VM measurements data in the network may be attacked before or after storage.
The first layer of the architecture constructs the semi-finished block after verifying the generated data packets, and the second layer performs PoW tasks on the semi-finished block to generate tamper-resistant metadata to ensure data integrity.
The system also includes an access control that can be updated to ensure that sensitive information cannot be accessed by the public.
{\it Mchain} separates the consensus process from the original blockchain and executes time-consuming PoW tasks in the background.
From users' view, not only can it guarantee security, but also reduce the packet waiting time.
This scheme can be improved by incorporating a consensus protocol that is more efficient than PoW.

Han {\it et al.} apply blockchain to medical data storage to build a decentralized, secure, tamper-resistant health information storage system in~\cite{han2018architecture}.
In order to improve the latency of data validation, the system utilizes a hybrid health blockchain, which combines the consortium and private blockchains.
The encrypted medical records are stored in the private blockchain, which is used as a database in medical institutions.
The consortium blockchain enables medical information sharing between nodes by storing medical data submitted from all participating medical institutions.
The node that wants to share its medical record can add its block to the consortium blockchain from the current private blockchain directly, thus achieving secure medical information storage, privacy protection, and medical data sharing.
However, the cost of deploying the proposed model is not considered in this article.

A blockchain-based data storage method is proposed to implement the secure domain name system in~\cite{liu2018data}.
This method can create multiple domain name service nodes in parallel, and the hash value of the file data is stored in the blockchain.
To securely store the log files, a platform based on blockchain in the cloud is presented in~\cite{kumar2018secure}  to achieve the integrity of the log files and logging process as well as the proof of non-repudiation.

\subsection{Trust-free Architecture}

The issues with trust in traditional data-driven networks can be efficiently addressed by blockchain technology in BDNs, which can build a trustable platform to share data and execute computations among different stakeholders and organizations.
In BDNs, users can feel safe to share transaction data with stranger nodes by leveraging the advantages of blockchain, which provides distributed and immutable ledger with efficient consensus protocol to ensure the trustfulness.

Huang {\it et al.} in~\cite{Huang_TLBN} propose a revocable chameleon hash (RCH) based on complexity assumptions and bilinear pairing for deriving a self-redactable blockchain (SRB) to enable an intelligent trust-layer for IoT.
Specifically, a collision could be found by RCH via ephemeral trapdoor and the SRB could enable the block content in the blockchain to be re-written and the redacted block hash to remain unchanged without suffering hard forks.

{\it BlockTDM} is a blockchain-based trusted data management scheme for mobile edge computing ~\cite{Ma_TDM}.
The core of {\it BlockTDM} is a configurable blockchain architecture, which consists of the edge device layer, blockchain network layer, edge nodes layer, and the cloud center layer.
The data gathered from the edge devices as well as their hash values are transmitted to the blockchain network layer for storage.
{\it BlockTDM} supports matrix-based multi-channel data segment and sensitive data isolation, by defining a channel matrix to perform data access, transfer and usage safely and effectively in an untrusted environment.
Moreover, users are capable of defining the data sensitivity, and {\it BlockTDM} can encrypt the transaction body, ({\it i.e.}, data payload) before saving the transaction in the blockchain system.
Using smart contract, conditional access is also implemented with decryption algorithms for users who want to access the protected blockchain and transaction data.
{\it BlockTDM} provides a versatile blockchain-based paradigm for trusted and tamper-proof data sharing and process.

Zhang {\it et al.} propose a blockchain-based trust scheme and elaborate a quality assurance application of blockchain in~\cite{Zhang_BTM} to enhance data security and partners' collaboration in smart manufacturing.
Each block records different kinds of digital data and information via asymmetric encryption, including task information, service information, identity data, transaction data, asset data, contract data, {\it et al.}
The block is verified and stored by all network nodes after broadcast, and a consensus would be achieved on the latest blockchain.
This scheme can build the trust of participants in the system and the trust between participants.
Its main challenges lie in the reliability of data source, the operating cost of the blockchain, and the deployment of smart contracts.

A blockchain-based decentralized trust management system for vehicular networks is designed in~\cite{Yang_TM_VN}.
The authors consider two types of adversary models including malicious users and damaged RSUs, and use blockchain to achieve decentralized trust management in the network.
Vehicles can validate the messages leveraging Bayesian inference model, and then generate and upload a rating result for each received message from other vehicles.
The RSUs calculate the trust value offsets of these rating results, and then upload the block containing these data to blockchain using a consensus mechanism combining PoW and PoS.
The more total value of offsets (stake) is in the block, the easier an RSU can find the nonce for the hash function.
In this way, all RSUs collaboratively maintain a reliable and trustful blockchain for the vehicular network.
However, how to jointly realize privacy protection and effective trust management is still a problem to be solved in such vehicular networks.
Another privacy-preserving anonymous reputation system (BARS) is proposed in~\cite{III5} to realize trustable VANETs.
Under the trust model of privacy protection managed by semi-trusted authorities, the blockchain enables transparency for the certification and revocation with the proofs of presence and absence.
All broadcast messages are recorded in ``Blockchain for Messages'' as permanent evidence to evaluate vehicle reputation, thus achieving an effective trust model in VANET.
Furthermore, a reputation evaluation algorithm is presented with both direct historical interactions and indirect opinions about vehicles to guarantee the reliability of messages.
Yang {\it et al.} propose a blockchain-based traffic event validation and trust verification mechanism ~\cite{Yang_BTEV}.
The authors assume that there is a Public Key Infrastructure (PKI)-based model in VANET and only consider events that can be verified by the vehicles or RSUs.
A proof-of-event (PoE) consensus algorithm is utilized by RSUs to detect passing vehicles within certain adjacent area in an incident when the collected data meets the corresponding threshold.
Meanwhile, the related information of incidents are recorded in the blockchain permanently.
The transactions on blockchain include two consecutive stages, namely, synchronizing the local blockchain and synchronizing the global blockchain.
In this way, the warning messages could be broadcast in an appropriate region and time periods.
The proposed mechanism can effectively record the correctness of traffic events, and provide traceable events with trust verification.

\section{Management and Control of Blockchain-empowered Data-driven Networks}
\label{sec:manage_survey}

In this section, we discuss DDN management and control supported by blockchain from the perspectives of access control, routing mechanisms, virtual resource management, resilient network and fault-tolerant mechanisms, as summarized in  Table~\ref{table:Manage&Ctrl}.

\begin{table*}[!th]
  \centering
  \caption{Management and Control of BDNs.}
  \label{table:Manage&Ctrl}
  \begin{tabular}{m{3.3cm}|m{1.5cm}<{\centering} m{8cm}}
  \hline \multicolumn{1}{c|}{Focus}  &  \multicolumn{1}{c}{Ref.}  &
  \multicolumn{1}{c}{Contributions}
  \\ \hline

 \multirow{8}{3.3cm}{Access Control}
  & \cite{Ugobame}   & {Use blockchain to manage access authority of large datasets. }
  \\ \cline{2-3}
  & \cite{Wang_ACBC, Tan_ACBC}   & {Propose an access control framework with integrated message encryption.}
  \\ \cline{2-3}
  & \cite{Ding_ABAC, Laurent} & {Present an attribute-based access control scheme for IoT.}
  \\ \cline{2-3}
  & \cite{Ma_ACBC, Paillisse}  & {Achieve access control in a cross-layer and cross-domain network.}
  \\ \cline{2-3}
  & \cite{meng2018intrusion, ujjan2019snort, alkadi2020deep}  & {Improve security of access control through blockchain-assisted intrusion detection.}
  \\ \hline

   \multirow{7}{3.3cm}{Routing Mechanisms}
  & \cite{Saad_Routing}  & { Propose a Blockchain-based secure BGP routing system.}
  \\ \cline{2-3}
  & \cite{li2019mitigating, ramezan2018blockchain, yang2019trusted}  & {Use blockchain to enhance the security and efficiency of routing in IoT and WSN.}
  \\ \cline{2-3}
  & \cite{dinginterchain, chen2017inter, Anlink}  &  {Design routing scheme to support cross-chain communication and routing.}
  \\ \cline{2-3}
  & \cite{ersoy2018transaction}  & {Use blockchain to combine incentive mechanism with smart routing.}
  \\ \hline

  \multirow{3}{3.3cm}{Virtual Resource Management}
  & \cite{rawat2019fusion, samaniego2017virtual, zhu2018edgechain}  & {Propose virtual resource placement scheme supported by blockchain at the edge of the network. }
  \\ \cline{2-3}
  & \cite{backman2017blockchain, li2019blockchain}  & {Design resource management and sharing mechanism in IoT based on blockchain.}
  \\ \hline

  \multirow{5}{3.3cm}{Resilient and Fault-tolerant Mechanisms}
   & \cite{zhou2018beekeeper, shi2018mpcstoken}  & {Study on the network fault-tolerance mechanism supported by blockchain. }
  \\ \cline{2-3}
  &  \cite{sagirlar2018autobotcatcher} & {Present a blockchain-based botnet detection architecture for IoT.}
  \\ \cline{2-3}
  & \cite{mylrea2017blockchain, liang2017towards}   & {Utilize blockchain to build distributed resilient network architecture.}
  \\ \hline

  \end{tabular}
  \end{table*}

\subsection{Access Control}

There have been many proposed approaches of blockchain-based access control designed for supporting different kinds of data-driven services.
In BDNs, there are many factors that should be considered in access control, including privacy protection, authentication, and resource allocation, with the target of sending and deriving data in a secure, efficient, and decentralized manner.

The blockchain-based access control system presented in  ~\cite{Ugobame} aims to effectively manage access authority of large datasets and protect against data breaches.
The Hyperledger fabric blockchain~\cite{cachin2016architecture} is designed to deploy blockchain identity-based access control (BIBAC) and blockchain role-based access control (BRBAC).
BIBAC supports access on a user-by-user basis, while BRBAC assigns users' roles to access specific assets on the blockchain.
Hyperledger modeling tool is adopted for the smart contract and transaction processing functions.
Though Hyperledger modeling raises questions about stability, the analysis indicates that ecosystem ensures data transparency and traceability for secure data sharing, auditabilty and data self-sovereignty for the owner.

Wang {\it et al.} in~\cite{Wang_ACBC} propose a secure cloud storage framework with access control by  Ethereum blockchain and ciphertext-policy attribute-based encryption (CP-ABE).
The blockchain is responsible for storing the publicly available information, achieving supervision function, and tracking the behavior of the data access.
The cipher-text data is stored by the data owner, and could be decrypted and used within the valid data-processing period via leveraging smart contracts.
When the attributes of data user satisfy the corresponding requirement and threshold, the data user could decrypt correctly and access the data before the valid data-processing period expires.
By using Ethereum smart contract, this framework can be combined with most CP-ABE algorithm to achieve decentralized fine-grained access control.

An access control architecture based on private blockchain is proposed in~\cite{Tan_ACBC} for information-centric networks (ICN), where the content is divided into $n$ original blocks, and encoded into $m~(m>n)$ blocks by xor-coding algorithm. The xor-based encoding/decoding scheme could encrypt and decrypt messages efficiently.
In order to meet the requirement that only authorized users can read and decrypt the protected content cached in ICN nodes, the authors make some security assumptions for the ICN environment.
Specifically, the content provider (CP) constructs a block containing the decryption information in a given sequence, and signs this block with its private key.
Then the CP publishes the blockchain and adds new blocks for updating.
The blocks are verified by users with the CP's public key via downloading the whole blockchain from ICN.
The users find their own information and decrypt it using their private key.
Unlike other strategies, there is no need to request decryption information from the CP, which makes it unnecessary for the CP to be always online and thus improves the robustness of the ICN.
Further security analysis is needed to prove the security of the framework in the presences of malicious nodes and attacks.

An attribute-based access control (ABAC) scheme for IoT is presented in~\cite{Ding_ABAC}.
It abstracts the roles or the identities into a set of attributes issued by the attribute authorities, which manage each of the attributes and distribute them to proper users.
The authors use consortium blockchain to record the attributes in a distributed manner to address the single point of failure and unauthorized data tampering problems.
The attribute authorities act as the serving nodes in a consortium blockchain and as the key generation center (KGC) when new IoT devices register with the system.
The devices could collect, transmit, process, and share the data in the system, and associate with the access policies according to their functional and secure requirements.
Note that the devices must prove their ownership of the corresponding attributes that satisfy the policy.
Thus, the devices are not responsible for transaction verification but have only read permission with the blockchain according to their attributes.
By reducing the overall computing and communication overheads and enhancing the system flexibility, this design is suitable for IoT scenarios.
Another blockchain-based ABAC approach is designed in~\cite{Laurent} to keep track of users' attributes, and manage the trust level with the attribute issuing entities (AIE) by administrators.
If the calculated trust level is higher than the corresponding threshold set by the devices, the attribute could be considered valid.
When all the required attributes are valid, the access request is granted.
Otherwise, the access request is denied.
This decentralized access control system solves the problem of lacking trust, and further research work need to consider the issues of information leakage and latency.

In~\cite{Ma_ACBC}, the authors propose a privacy-oriented blockchain-based distributed hierarchical access control framework, which consists of the cloud layer, fog layer, and device layer.
The edge network provides the access function for the devices.
In the fog layer, the security access manager is responsible for recording and verifying transactions that include key management information.
The cloud layer with multi-blockchains helps to achieve interconnection and traceability among blockchains.
The cloud layer also stores the encrypted data generated by all kinds of mobile devices, and the data can be accessed directly using the encryption key.
The network is divided into different side-blockchains to speed up the verification and save the storage space.
This framework achieves decentralized, fine-grained auditability and scalability requirements. However, blockchain technology is not fully utilized.

Paillisse {\it et al.} implement and evaluate a three-layer architecture in~\cite{Paillisse}, which supports distributed access control based on permissioned blockchain in cross-domain communications.
Private blockchains and BFT protocols are applied to this structure, so that moderate storage space can be used to store thousands of access strategies.
Authorized user could access network resources, and routers could determine the authority of the users.
The Locator/ID Separation Protocol (LISP) is adopted to support the communications of control and data planes.
The control plane stores specific access policies and updates them via the  Hyperledger, which brings secure and tamper-proof features.

Besides, blockchain-assisted intrusion detection plays an important role in identifying possible threats and achieving secure access control in DDNs~\cite{meng2018intrusion}.
To enhance the detection capabilities of the intrusion detection system (IDS), Ujjan  {\it et al.} in~\cite{ujjan2019snort} combines blockchain with the collaborative intrusion detection network (CIDN) in SDN to achieve trust-based communication among IDS nodes.
Not only can each node monitor and identify malicious traffic in the network, but also share the Snort signature rule set with neighboring nodes.
This CIDN model provides protective measures against internal attacks, enabling accurate detection of some typical common attacks such as DDoS.
Unsupervised deep learning methods can further improve the performance of such frameworks.
A deep blockchain framework (DBF) is proposed in~\cite{alkadi2020deep}, which uses a Bidirectional Long Short-Term Memory (BiLSTM) algorithm to discover network attacks from network data migration in cloud systems.
This intrusion detection algorithm was evaluated on the UNSW-NB15 and BoT-IoT data sets.
The results show that the proposed framework can achieve simple, secure, and transparent intrusion detection between clouds, supporting users and cloud providers for secure access and data migration.
Note that although blockchain-based IDS can achieve security and privacy in access control, in terms of system latency and scalability, blockchain-based IDSs should be further tested and improved in the DDNs.

\subsection{Routing Mechanisms}

BDNs can provide services to collect, process, and store data and information through networks with single or multiple hop relays.
By taking advantage of blockchain as a trusted, decentralized, self-organizing ledger system, effective routing mechanisms have been proposed by many researchers.

A blockchain-based secure Border Gateway Protocol (BGP) routing system named {\it RouteChain} is proposed in~\cite{Saad_Routing}, to prevent BGP hijacking and maintain a consistent view of the Internet routing paths.
Autonomous systems (ASs) are grouped according to the geographical proximity, and then a bi-hierachical blockchain-based model is built to detect misbehavior over the Internet.
Each transaction contains BGP announcements that are exchanged among peers, and a PoA-based consensus protocol called {\it Clique} is used to achieve consensus among these ASs.
Since the propagation and verification of transactions between ASs takes some time, {\it RouteChain} cannot prevent all ASs from being attacked completely.
In addition, grouping ASs according to the ideal situation may not work well in reality.
Despite the above defects, this architecture can still reduce BGP attacks in the network effectively.

A blockchain-based distributed reputation management system is presented in ~\cite{li2019mitigating} to protect secure routing in IoT networks.
Considering the behavior of malicious routers refusing to provide routing services by discarding data packets, the system rates the reputation of each router and stores it in the blockchain in a decentralized and immutable manner.
The reputation results are used to evaluate the trustworthiness of each router to protect the routing mechanism from misbehavior.
To reduce the complexity of blockchain in IoT networks, a group mining scheme rather than PoW-based consensus protocol is designed to achieve consensus among IoT nodes.
The proposed scheme can realize reputation management of the IoT routing process against  selfish nodes, and has stable convergence under different network scales.
Further analysis needs to be done to evaluate the security performance of this routing system.
Targeting an untrusted IoT environment,
Ramezan {\it et al.} propose a blockchain-based contractual routing (BCR) protocol in~\cite{ramezan2018blockchain}.
Each source IoT device implements a smart contract on the blockchain to update topology.
The routing algorithm coded in the smart contract finds the proper route from a source IoT device to a destination device or gateway, and the effectiveness of the route could be guaranteed by any intermediary device.
In order to enhance routing security and efficiency for wireless sensor networks (WSNs), reinforcement learning is adopted in~\cite{yang2019trusted} to propose a blockchain-based trusted routing scheme.
The authors assume that the blockchain network is trusted and the nodes in WSN can be static or dynamic, which use the proposed blockchain platform to provide dynamic and trusted routing.
Each transaction contains routing information. The routing packets, confirmed by the verification nodes, are encapsulated into blocks and stored in the blockchain, which guarantees that these routing packets are traceable and tamper-proof.
Then, a reinforcement learning model is utilized to adaptively select the best route to destination nodes.
The system can effectively suppress attacks by malicious nodes while guaranteeing lower latency and better throughput performance.

Apart from these blockchain-based routing solutions, some researchers consider adding routing function to make selected blockchain nodes transmit requests between different blockchain networks.
{\it Interchain} is proposed with a handshaking method to complete asset transfer for cross communications between blockchains~\cite{dinginterchain}.
However, it does not consider any consensus algorithm in the framework.
Chen {\it et al.} in~\cite{chen2017inter} introduce a private token-based inter-blockchain communication scheme to provide  crossover communications between different blockchains without any intermediary.
The authors utilize a routing algorithm and PBFT as the consensus algorithm. The main limitation of this work is that the proposed method negatively affects the system throughput.

{\it Anlink} blockchain~\cite{Anlink} is an enterprise blockchain architecture that uses an inter blockchain communication protocol to connect multiple blockchains and enable cross-chain communications.
The proposed architecture is composed of {\it Ann-Router}, {\it AnnChain}, as well as other blockchain-based systems.
These components are divided into four participants, namely, the validator, surveillant, nominator, and connector.
Ann-Router is a blockchain router to dynamically maintain all the related information registered on sub-chains and link sub-chains, and enable multiple blockchains to communicate with each other in the network.
Delegated stake-PBFT is used as the consensus protocol.
In addition to meeting the communication needs between blockchains, enterprise blockchain systems also need to meet the regulatory requirements, provide privacy protection, and support heavy transaction loads.

Ersoy {\it et al.} in~\cite{ersoy2018transaction} combine an incentive mechanism with smart routing to reduce the communication and storage cost.
Each participating node in the propagation of a transaction could receive a share of the transaction fee.
They analyze the sufficient and necessary conditions that encourage the spread of messages as well as to discourage the network from introducing Sybil nodes, and design a routing mechanism considering a first-leader-then-block type consensus protocols where the round leader who creates the block is known in advance.
The proposed routing mechanism reduces the propagation of redundant communication while encouraging nodes to propagate messages.

\subsection{Virtual Resource Management}

Wireless network virtualization is considered as a promising technology to enable sharing of physical infrastructure and networking slices for enhancing network capacity, coverage, and wireless security.
Some pioneer work have been carried out in BDNs.

An architecture leveraging SDN, edge computing, and blockchain is proposed in~\cite{rawat2019fusion} to enable wireless network operators to utilize their resources efficiently and securely.
Implemented as an overlay architecture on existing networks, the key components include SDN controller, network aggregator, primary wireless resource owners (PWROs), virtual wireless network operators (VWNOs), edge computing component, and blockchain platform with its manager.
Specifically, the sublease and release of networking slices between PWROs and VWNOs could apply blockchain to offer traceability and auditability so as to prevent double-spending of the same networking slice in the same period and location for wireless virtualization.
SDN controllers, which act as blockchain managers, keep their own private keys, and use public keys while subleasing (or releasing) networking slices to VWNOs (or from VWNOs to PWROs).
Therefore, the sharing framework in~\cite{rawat2019fusion} allows users to move/switch between virtual networks while maintaining a secure connection.
The blockchain guarantees that it is impossible to generate malicious transactions that sublease wireless resources with authorization; however, open issues including large storage space and long consensus delay still need to be addressed.

Samaniego {\it et al.} in~\cite{samaniego2017virtual} introduce a virtualization of IoT components (virtual resources) at the edge of  IoT networks.
Permission-based blockchain protocols are introduced including blockchain as a service (BaaS) in the cloud and a private multi-chain in a fog network, to enable provisioning of virtual resources directly on edge devices.
Only registered users can access the blocks in the chain to write and read configurations, thereby managing the virtual resource provisioning and multi-tenant access in a secure manner.
Considering the latency and bandwidth consumption, the fog layer implements a multi-chain for hosting virtual resources and only transfers useful information to the cloud.
The evaluation of data is also performed in the fog layer to provide a time-effective decision-making process.

The authors in~\cite {backman2017blockchain} present the concept of using blockchain as the ledger for network slice leasing, and analyze some use cases for future industrial IoT networks.
A 5G Network Slice Broker is introduced into blockchain in this work, enabling the factories to obtain the required slices automatically and dynamically to achieve efficient operation on-demand.
Blockchain is applied to manage virtual network slice trading for secure network operation, and enable new functionalities, such as spectrum management, data processing, and network infrastructure as a service.

A blockchain-based workflow management system (BCWMS) is proposed in~\cite{li2019blockchain} to share heterogeneous logistics resources among different customers to satisfy different operation logic.
To simplify the process of systems and resource association for IoT devices, the virtual resource gateway (VRG) is introduced to enable resources management with different granularity via specific gateways and virtual links.
Also, a resource blockchain is designed to guarantee data reliability and the accuracy of front-line resource usage data for enabling customer decisions.

{\it EdgeChain} is designed in~\cite{zhu2018edgechain} to build a decentralized platform for mobile edge application without any trusted third party.
The authors use stochastic programming for multiple service providers to make mobile edge application placement decisions.
The cost is modeled by jointly considering the edge hosts, latency, and service chaining.
The blockchain is used to store all placement transactions, including global resource availability, allocation, and consumption information, which are traceable by every mobile edge service provider and application vendor who consumes virtual resources.
However, this work does not take into consideration about the users' behavior, which may have an impact on system cost and resource deployment.

\subsection{Resilient and Fault-tolerant Mechanisms}

In BDNs, fault or malicious activities should be readily identified and recovered to enable a decentralized, resilient and fault-tolerant system.

{\it BeeKeeper} is proposed in~\cite{zhou2018beekeeper} as a fault-tolerant blockchain-based IoT service system, which implements a threshold secure multi-party computing (TSMPC) protocol. Servers perform homomorphic computations on shares and generate responses to help users processing encrypted data.
Since shares and responses are verifiable by leveraging blockchain, malicious nodes can be easily detected. However, the main limitation of {\it BeeKeeper} is that the number of active and honest servers should be more than a threshold to keep the protocol work effectively.

The authors in~\cite{shi2018mpcstoken} propose a fault-tolerant incentivisation mechanism based on payment strategies conditional on task execution results for distributed mobile peer-to-peer crowd services (MPCS) systems.
Moreover, smart contract {\it MPCSToken} is designed to facilitate service auction, task execution and payment settlement process.
{\it MPCSToken} contract is implemented on Ethereum blockchain to build a micropayment mechanism based on blockchain payment channels to avoid bulky payments and mitigate payment settlement risks.
The system can improve the utility of participants effectively, and can also run cost-effectively on the Ethereum blockchain.

A blockchain-based botnet detection architecture for IoT, called {\it AutoBotCatcher}, is presented by Sagirlar {\it et al.} in~\cite{sagirlar2018autobotcatcher} to analyze communities of IoT devices and detect botnets.
A permissioned BFT blockchain is utilized to store the traffic flows of the IoT devices.
A set of pre-identified parties collect and audit these traffic flows as blockchain transactions to perform botnet detection collaboratively without any trusted third party.
Specifically, there are two main actors, namely agents and block generators.
Agents could monitor IoT network traffic flows and send collected traffic information as blockchain transactions.
Block generators ({\it i.e.}, the powerful trusted full node in the IoT domain) aim at modeling mutual contact information of IoT devices and generating mutual-contact graph, which is then exploited to detect the logical communities.
Based on these, {\it AutoBotCatcher} can perform dynamic and collaborative botnet detection on large networks, but its robustness and the ability to deal with internal threats still require further evaluation.

Mylrea {\it et al.} in~\cite{mylrea2017blockchain} investigate the application of blockchain and smart contract, which provides anatomically verifiable cryptographic signed distributed ledger to build decentralized, resilient, and smart energy grids as well as energy trading platforms.
The authors focus on the technical characteristics of blockchain (security, scalability and speed), thus apply the blockchain to provide an innovative trading platform where producers and consumers can exchange their residual energy or demand flexibly.
Blockchain is utilized to verify time, user, and transaction data, and protect these data with an immutable cryptographic signed ledger to improve the trustworthiness, integrity, and resilience of energy delivery systems.
Moreover, energy customers can also verify data from other entities by leveraging blockchain to creat a distributed trust mechanism.

A trusted and resilient blockchain-based architecture for IoT is designed in ~\cite{liang2017towards} to support data integrity auditing and enable better resilience to improve the system scalability.
It considers a system in which drones are deployed to provide better connectivity among IoT devices.
Specifically, the hashed data records collected from drones are stored in the blockchain network, and each blockchain receipt for each data record is generated and stored in the cloud considering the limitations of battery and processing capability of drones while guaranteeing data security.
To improve robustness, a full copy of the entire distributed ledger is stored on every distributed node in the blockchain network.
The system can improve the network reliability and security, as well as collect data in real time to provide data guarantee for drone control.

\section{Challenges and Future Directions for Blockchain-empowered Data-driven Networks}
\label{sec:challenges}

In this section, we discuss the challenges and future directions for BDNs in terms of security and privacy, scalability, network management, and resource management. A summary of the references for these issues can be found in the online supplemental file. Furthermore, standardization activities related to BDNs have been gaining momentum, some of which are also presented in this section.

\subsection{Security and Privacy Issues}

Although blockchain technology could provide a promising solution for enhancing security and privacy in a variety of applications, BDNs, which integrates blockchain with DDNs, may encounter new security challenges, such as selfish mining attacks~\cite{eyal2018majority}, balance attacks~\cite{natoli2016balance}, BGP hijacking attacks~\cite{apostolaki2017hijacking}, and eclipse attacks~\cite{singh2006eclipse}.
In ~\cite{wilkinson2014storj}, blockchain is used as a common record for all user encryption metadata.
An attacker can pretend to be a legitimate node to read encrypted metadata, and then attempt to decode it through an offline brute-force attack. Therefore, appropriate and effective security measures must be developed for BDNs.
One of the main advantages of blockchain is the provision of pseudo-user anonymity, which is critical to safety.
For most of the methods discussed above, the transaction data records can reveal the user's identity by looking into the public key, the ACL, or the provenance data.
Therefore, by analyzing the data stored in a blockchain, the users' behavioral activities can be tracked, and their true identities can be exposed.
To solve this problem, further research is needed to provide a completely anonymous approach that meets the security requirements.
Another issue is the possibility of quantum attacks.
The encryption of blockchain will no longer be secure when practical quantum computers are realized~\cite{Aggarwal2017Quantum}.
Therefore, new ``quantum-safe'' encryption mechanisms will need to be developed for future BDNs.

Machine learning can be adopted to support BDNs, because ubiquitous blockchain nodes with machine learning algorithm can react to network status and users' requirements in a decentralized and timely manner, which not only improves  robustness but also reduces latency.
The application of  ML in BDNs can detect malicious behavior on the blockchain by deploying the trained models and algorithms.
In this way, ML could assist blockchain systems in identifying and preventing theft, fraud, and illicit transactions on the chain.
Furthermore, a blockchain can store large-scale data in a secure and tamper-resistant manner, ensuring data confidentiality and auditability of the collaborative training process and the trained ML model via cryptographic techniques.

The world is experiencing a data explosion in both amount and diversity.
Networking data is continuously generated by users and network entities.
The approaches to store, organize, and process these data are being actively explored.
The combination of  blockchain and big data in the network can reduce operating costs and data storage risks in BDNs.
Towards this direction, Yue {\it et al.} in~\cite{yue2017big} propose a blockchain-based big data sharing model to improve data credibility and ensure secure data circulation.
Also, to enhance the security of big data platforms, an access control framework is presented in~\cite{es2017blockchain} by relying on blockchain to  ensure proper implementation of optimal access in a decentralized environment without central authority and administrators.
At the same time, the historical data of the network can quickly identify and prevent malicious authentication and authorization behaviors through data mining and real-time analyzing, thereby ensuring the cybersecurity for future BDNs.
However, it may lead to huge data redundancy.
Therefore, how to perform data off-chain operation on the premise of ensuring privacy and integrity is a problem that should be further investigated.

\subsection{Scalability Issues}

When blockchain technology is applied to future DDNs, scalability is another important problem that needs to be addressed. For example, the Bitcoin blockchain requires more than 100 GB of storage and can handle only 7 transactions per second on average.
If we increase the volume of each block ({\it i.e.}, more transactions are packed together), its throughput could increase, but generating and propagating blocks would take a longer time.
On the other hand, reducing the block interval time can increase throughput, but this can also lead to stale blocks that have no contribution to the main chain and reduce the security.
Thus, designing blockchain for BDNs requires a trade-off between scalability and security~\cite{ShardingBlockchains}.

To enhance scalability of a BDN, the designed blockchain must also ensure its security. In order to solve the problem of storage, Bruce {\it et al.} in~\cite{miniblockchain} proposed a novel blockchain structure, which uses a database called {\it account tree} and deletes the old transaction records instead of holding all addresses.
Then, the nodes only need the latest part of the blockchain to synchronize with the network, saving storage space in the blockchain.
The low storage nodes proposed in~\cite{Perard2018Erasure} store linearly combined fragments of each block by using erasure codes.
These nodes can recover data by downloading fragments from other nodes and performing inverse linear operations.
This technology ensures that any node in the blockchain can be easily reconstructed from such nodes with the data integrity guaranteed.
However, compatibility of this technique with other networks still needs more investigation.

In terms of redesigning blockchain, the work in~\cite{Dorri2017Towards} proposes a theoretical lightweight blockchain structure, which can handle transactions without incurring additional delays.
However, without a comprehensive analysis of consensus and security, it is unclear whether it can withstand attacks from corrupted nodes in the network.
Poon {\it et al.} in~\cite{Plasma} propose a solution for scalable computing on blockchains named Plasma, which is scalable to a significant amount of state updates per second.
Plasma divides all blockchain calculations into a set of MapReduce functions.
Each sidechain is implemented with a smart contract, and the root chain only needs to handle a small amount of commitment from the sidechain.
The scalability of the root chain can thus be enhanced.
However, Plasma can only guarantee security and scalability at the same time when the withdrawal delays is fixed; otherwise its security level may reduce.

To address the scalability issues, ML techniques can benefit the blockchain in making better decisions, such as offering more efficient data sharding or pruning solutions.
Leveraging ML techniques, blockchain applications can support predictive analytics to ensure the requirements for resources to be accurately met and to improve the efficiency of blockchain operation.
Specifically, leveraging the training data, ML-based mining algorithms may manage tasks in a more intelligent manner rather than adopting the brute-force approach.
Since ML algorithms can predict and speedily calculate data, it would also provide a feasible way for miners to select more important transactions to perform.
ML techniques can also enable more efficient off-chain solutions or adjust the block size dynamically to make the blockchain-based system more scalable.

\subsection{Network Management Issues}

Intelligent network management enables the introduction of autonomous mechanisms into the network to perform tasks such as planning, configuration, management, optimization, and self-repair.
Applying blockchain to the DDN and correlating functions by using smart contracts that can be executed automatically is a good way to implement autonomous functions in intelligent management.
A new intelligent network architecture is proposed in~\cite{Maksymyuk2019}, which uses smart contracts to handle the relationship between operators and users, thereby simplifying spectrum and infrastructure sharing.
At the same time, the consensus protocol can also make the network highly resilient and resistant to DoS flooding.
ADEPT~\cite{panikkar2015adept} enables the devices connected to it to communicate in a secure and efficient manner in the IoT network, while supporting complex business logic.
It also integrates multiple protocols to achieve an autonomously coordinated, secure, and decentralized network.

However, because of distributed data storage used in blockchain, each node maintains a complete copy of the data.
Once a new transaction is added to the block, all copies of the data should be updated synchronously. This frequent data updates may increase the difficulty of blockchain-based network management.
Neudecker {\it et al.} in~\cite{neudecker2018network} identify five requirements for the blockchain network layer: performance, low cost of participation, anonymity, DoS resistance, and topology information hiding.
The network layer needs to be weighed and optimized in these five areas.
However, most of these requirements are not quantitatively described.
Therefore, the design of the blockchain network should be further explored, and the data in the network should be used to improve the performance of blockchain.

Smart contracts in the blockchain can enable consensus-based coordination and verification in the network in a manner beneficial to network management, but there are some issues that require further investigations.
One of the problems is that in the basic construction of decentralized autonomous management, it is difficult to fix bugs once the system is in operation~\cite{chohan2017decentralized}.
Another problem is the participant management problem in the decentralized autonomous organization.
This involves how to manage participants of the consensus mechanism, and how to develop incentives and accountability mechanisms.
Important details and new capabilities need to be considered to promote an autonomous, secure, transparent and more efficient network management.

Applying edge computing to BDNs can achieve reliable access and control for edge-distributed networks, thereby providing large-scale network control and management services.
The storage resources on edge computing devices can be used for content caching, with BDNs maintaining a credit system to support the business model.
One example of this approach is the modular consortium architecture based on blockchain and IPFS software stack presented in~\cite{ali2017iot}, which contains private side-chains and a consortium blockchain.
The private side-chains maintain logs of data operations that occur within a private IoT network and the consortium blockchain connects edge servers, thus achieving security and privacy of IoT data.
In addition, the application of big data can improve the automatic optimization of the future networks, enabling BDNs to realize self-adjustment and intelligent networking.

\subsection{Resource Management Issues}

Application of blockchain in DDN could make the resource management problem more critical. A decentralized cloud resource management framework based on blockchain techniques is presented in~\cite{xu2017intelligent} to save energy significantly.
This framework schedules requests without relying on a centralized scheduler in the cloud.
The decentralization of the blockchain ensures that the failure of a data center does not affect the system's resource management, which gives the framework excellent robustness.

Another issue is the resources consumed by a blockchain.
In order to reach decentralized consensus among nodes with poor synchronization, some existing consensus protocols in the blockchain networks ({\it e.g.}, PoW protocols) will consume a huge amount of computing and energy resources.
To reduce the resource consumption of the blockchain, some other alternative PoX consensus protocols have been proposed in permissionless blockchain systems~\cite{ball2017proofs, zhang2017rem, park2015spacecoin}.
These PoX solutions address two main goals, namely, incentivizing the provision of useful resources and improving performance~\cite{wang2019survey}.
However, the resources required by these consensus protocols have not been extensively analyzed and quantified.

Edge computing provides a low-latency and distributed computational task offloading platform for end nodes with limited resources.
Edge computing decomposes the large-scale services conventionally handled by the  data centers (i.e., in the cloud), dividing them into smaller parts and distributing them to the edge nodes for processing, thereby speeding up data processing and reducing latency.
This aligns very well with the decentralized machine learning solution that we mentioned above for future BDNs.
An example is the blockchain-based computation offloading scheme in~\cite{xu2019become} that protects data integrity in task offloading while optimizing resource allocation using a  non-dominated sorting genetic algorithm.
A deep learning method to optimize an auction-based resource allocation scheme for edge servers is presented in~\cite{luong2018optimal}. However, it only considers an individual edge server. A feasible approach for interactions between multiple edge servers to optimize resource utilization while satisfying latency and other performance constraints should be developed to support BDNs.
\\
\bigskip
\subsection{Standardization Activities Related to Blockchain-empowered Data-driven Networks}
In order to speed up the development, implementation and adoption of BDNs, standardization efforts are quite crucial to ascertain interoperability among network entities, user terminals, and various applications. While DDNs have attracted wide attention from researchers, their standardization work is still in the early stage.
ITU-T has recently issued the Y.3650 series to discuss the use-cases and application scenarios for big-data driven networking~\cite{ITU}.
The market and related industry are quite fragmented, and a widely acknowledged and commonly accepted framework covering different technologies and platforms is still lacking.
Meanwhile, standards for blockchains and DLT  have started to emerge since late 2017, and
several organizations have intensified their efforts to develop various standards~\cite{lima2018developing}.
The IEEE P2418 series, which focuses on generic architectures, interoperability, and vertical-industry standards for blockchain/DLT is under development.
ISO/TC307 is an international technical committee that has been very active in setting global standards in wide areas including use cases, smart contracts, security and privacy, governance, and interoperability between blockchain applications.
ITU-T Focus Group on Application of Distributed Ledger Technology (FG DLT) has been created to analyze the standardization demands of applications and services based on DLT.
The Enterprise Ethereum Alliance (EEA) has more than 500 members worldwide and is considered as one of the most active industry alliances.
It focuses on development of technical specifications and certification of Ethereum for enterprises.
The W3C blockchain community group has been working on new technologies and use cases related to blockchain.

Although there are many global standardization initiatives ongoing for DDNs and blockchain technologies, standardization efforts for BDNs need to be launched to support the vigorous development of an industry ecosystem.
Industry has an urgent need to standardize some aspects of BDNs. The application programming interface (API) formats should be unified among different manufacturers and service providers, so that data-driven services may bloom and flourish. The APIs between data engine and block engine need to be standardized to support the information exchange among various modules. Meanwhile, in order to support efficient data transfer across network operators, the processing method of the data ({\it i.e.} the label for sensitive data, authentication and authorization policies) calls for corresponding standards as well.
The joint efforts from interest groups, industry-driven forums, standards-development organizations, together with academic experts, will be needed to push the standardization and widespread adoption of BDNs.

\section{Conclusion}
\label{sec:conclusion}
The current generation computing networks are rapidly evolving towards future DDNs connecting a massive number of heterogeneous users and devices, generating and consuming a huge amount of data that require processing at strategic locations within the networks, all for the purpose of enabling ubiquitous intelligent services.
In this article, we have reviewed the fundamentals of the cutting-edge blockchain technologies and how they are proposed to be applied in computer networks. We have further analyzed the challenges of future DDNs and discussed how blockchain can help address many of these challenges.
We have systematically surveyed and categorized the pioneer research works of using blockchain in networks to motivate blockchain as a solution to empower future DDNs.
We have highlighted the specific advantages of BDNs, made possible by the decentralization of blockchain that is reshaping the conventional paradigms of running complex systems.
We have presented a framework for BDN, and discussed how decentralization can be beneficially applied to different layers ({\it i.e.}, application layer, network layer, and link layer) in various dimensions ({\it i.e.}, data plan and control plan) for multiple purposes ({\it i.e.}, security control, privacy protection, robustness enhancement) with different kinds of resources ({\it i.e.}, computational resource, bandwidth/communication resource, and storage resource).
We have also identified a number of research challenges, and presented the broader perspectives on how BDN might evolve in the future to incorporate AI, big data and edge computing.
We hope this article will provide impetus for a substantial increase in research towards the design and implementation of BDNs, as well as the enhancement of existing services and development of new services based on BDNs.

  \bibliographystyle{unsrt}
  \bibliography{ACM_Survey}

\begin{thebibliography}{100}

\bibitem{cisco2018cisco}
VNI Cisco.
\newblock Cisco visual networking index: Forecast and trends, 2017--2022.
\newblock {\em White Paper}, 1, 2019.

\bibitem{fang2019data}
Chao Fang, Song Guo, Zhuwei Wang, Huawei Huang, Haipeng Yao, and Yunjie Liu.
\newblock Data-driven intelligent future network: Architecture, use cases, and
  challenges.
\newblock {\em IEEE Communications Magazine}, 57(7):34--40, 2019.

\bibitem{ITU}
ITU-T Y.3650.
\newblock Framework of big-data-driven networking, 2018.

\bibitem{ref2}
Weichao Gao, William~G Hatcher, and Wei Yu.
\newblock A survey of blockchain: Techniques, applications, and challenges.
\newblock {\em 27th International Conference on Computer Communication and
  Networks}, pages 1--11, Jul. 2018.

\bibitem{ref3}
Satoshi Nakamoto.
\newblock Bitcoin: A peer-to-peer electronic cash system.
\newblock \url{http://bitcoin.org/bitcoin.pdf}.
\newblock Accessed May 4, 2019.

\bibitem{ref7}
Weisong Shi, Jie Cao, Quan Zhang, Youhuizi Li, and Lanyu Xu.
\newblock Edge computing: Vision and challenges.
\newblock {\em IEEE Internet of Things Journal}, 3(5):637--646, 2016.

\bibitem{ref8}
Wei Yu, Fan Liang, Xiaofei He, William~Grant Hatcher, Chao Lu, Jie Lin, and
  Xinyu Yang.
\newblock A survey on the edge computing for the internet of things.
\newblock {\em IEEE access}, 6:6900--6919, 2017.

\bibitem{ref6}
Fan Liang, Wei Yu, Dou An, Qingyu Yang, Xinwen Fu, and Wei Zhao.
\newblock A survey on big data market: Pricing, trading and protection.
\newblock {\em IEEE Access}, 6:15132--15154, 2018.

\bibitem{ref4}
John~A Stankovic.
\newblock Research directions for the internet of things.
\newblock {\em IEEE Internet of Things Journal}, 1(1):3--9, 2014.

\bibitem{ref5}
Jie Lin, Wei Yu, Nan Zhang, Xinyu Yang, Hanlin Zhang, and Wei Zhao.
\newblock A survey on internet of things: Architecture, enabling technologies,
  security and privacy, and applications.
\newblock {\em IEEE Internet of Things Journal}, 4(5):1125--1142, 2017.

\bibitem{yin2014big}
Hao Yin, Yong Jiang, Chuang Lin, Yan Luo, and Yunjie Liu.
\newblock Big data: Transforming the design philosophy of future internet.
\newblock {\em IEEE Network}, 28(4):14--19, 2014.

\bibitem{yao2016novel}
Haipeng Yao, Chao Qiu, Chao Fang, Xu~Chen, and F~Richard Yu.
\newblock A novel framework of data-driven networking.
\newblock {\em IEEE Access}, 4:9066--9072, 2016.

\bibitem{wang2016data}
Ying Wang, Peilong Li, Lei Jiao, Zhou Su, Nan Cheng, Xuemin~Sherman Shen, and
  Ping Zhang.
\newblock A data-driven architecture for personalized qoe management in 5g
  wireless networks.
\newblock {\em IEEE Wireless Communications}, 24(1):102--110, 2016.

\bibitem{han2017big}
Shuangfeng Han, I~Chih-Lin, Gang Li, Sen Wang, and Qi~Sun.
\newblock Big data enabled mobile network design for 5g and beyond.
\newblock {\em IEEE Communications Magazine}, 55(9):150--157, 2017.

\bibitem{cui2016big}
Laizhong Cui, F~Richard Yu, and Qiao Yan.
\newblock When big data meets software-defined networking: Sdn for big data and
  big data for sdn.
\newblock {\em IEEE network}, 30(1):58--65, 2016.

\bibitem{zheng2016big}
Kan Zheng, Zhe Yang, Kuan Zhang, Periklis Chatzimisios, Kan Yang, and Wei
  Xiang.
\newblock Big data-driven optimization for mobile networks toward 5g.
\newblock {\em IEEE network}, 30(1):44--51, 2016.

\bibitem{chen2018data}
Min Chen, Yongfeng Qian, Yixue Hao, Yong Li, and Jeungeun Song.
\newblock Data-driven computing and caching in 5g networks: Architecture and
  delay analysis.
\newblock {\em IEEE Wireless Communications}, 25(1):70--75, 2018.

\bibitem{cheng2018big}
Nan Cheng, Feng Lyu, Jiayin Chen, Wenchao Xu, Haibo Zhou, Shan Zhang, and
  Xuemin~Sherman Shen.
\newblock Big data driven vehicular networks.
\newblock {\em IEEE Network}, 32(6):160--167, 2018.

\bibitem{sammarco2019unsupervised}
Matteo Sammarco, Miguel Elias~Mitre Campista, Marcin Detyniecki, Tahiry
  Razafindralambo, and Marcelo~Dias de~Amorim.
\newblock Unsupervised detection of adversarial collaboration in data-driven
  networking.
\newblock In {\em 2019 10th International Conference on Networks of the Future
  (NoF)}, pages 1--8. IEEE, 2019.

\bibitem{astaras2019deep}
Stefanos Astaras, Sofoklis Efremidis, Angela-Maria Despotopoulou, John
  Soldatos, and Nikos Kefalakis.
\newblock Deep learning analytics for iot security over a configurable bigdata
  platform: Data-driven iot systems.
\newblock In {\em 2019 22nd International Symposium on Wireless Personal
  Multimedia Communications (WPMC)}, pages 1--6. IEEE, 2019.

\bibitem{huang2017data}
Haojun Huang, Hao Yin, Geyong Min, Hongbo Jiang, Junbao Zhang, and Yulei Wu.
\newblock Data-driven information plane in software-defined networking.
\newblock {\em IEEE Communications Magazine}, 55(6):218--224, 2017.

\bibitem{chih2017big}
I~Chih-Lin, Qi~Sun, Zhiming Liu, Siming Zhang, and Shuangfeng Han.
\newblock The big-data-driven intelligent wireless network: Architecture, use
  cases, solutions, and future trends.
\newblock {\em IEEE vehicular technology magazine}, 12(4):20--29, 2017.

\bibitem{ma2020survey}
Bo~Ma, Weisi Guo, and Jie Zhang.
\newblock A survey of online data-driven proactive 5g network optimisation
  using machine learning.
\newblock {\em IEEE Access}, 8:35606--35637, 2020.

\bibitem{zhang2020enabling}
Chaofeng Zhang, Mianxiong Dong, and Kaoru Ota.
\newblock Enabling computational intelligence for green internet of things:
  Data-driven adaptation in lpwa networking.
\newblock {\em IEEE Computational Intelligence Magazine}, 15(1):32--43, 2020.

\bibitem{uriarte2018blockchain}
Rafael~Brundo Uriarte and Rocco De~Nicola.
\newblock Blockchain-based decentralized cloud/fog solutions: Challenges,
  opportunities, and standards.
\newblock {\em IEEE Communications Standards Magazine}, 2(3):22--28, 2018.

\bibitem{yang2019integrated}
Ruizhe Yang, F~Richard Yu, Pengbo Si, Zhaoxin Yang, and Yanhua Zhang.
\newblock Integrated blockchain and edge computing systems: A survey, some
  research issues and challenges.
\newblock {\em IEEE Communications Surveys \& Tutorials}, 21(2):1508--1532,
  2019.

\bibitem{salah2019blockchain}
Khaled Salah, M~Habib~Ur Rehman, Nishara Nizamuddin, and Ala Al-Fuqaha.
\newblock Blockchain for ai: Review and open research challenges.
\newblock {\em IEEE Access}, 7:10127--10149, 2019.

\bibitem{ali2018applications}
Muhammad~Salek Ali, Massimo Vecchio, Miguel Pincheira, Koustabh Dolui, Fabio
  Antonelli, and Mubashir~Husain Rehmani.
\newblock Applications of blockchains in the internet of things: A
  comprehensive survey.
\newblock {\em IEEE Communications Surveys \& Tutorials}, 21(2):1676--1717,
  2018.

\bibitem{Zhu2019ACM}
Qingyi Zhu, Seng~W. Loke, Rolando Trujillo-Rasua, Frank Jiang, and Yong Xiang.
\newblock Applications of distributed ledger technologies to the internet of
  things: A survey.
\newblock {\em ACM Comput. Surv.}, 52(6), November 2019.

\bibitem{khan2018iot}
Minhaj~Ahmad Khan and Khaled Salah.
\newblock Iot security: Review, blockchain solutions, and open challenges.
\newblock {\em Future Generation Computer Systems}, 82:395--411, 2018.

\bibitem{zhu2018identity}
Xiaoyang Zhu and Youakim Badr.
\newblock Identity management systems for the internet of things: A survey
  towards blockchain solutions.
\newblock {\em Sensors}, 18(12):4215, 2018.

\bibitem{pohrmen2019blockchain}
Fabiola~Hazel Pohrmen, Rohit~Kumar Das, and Goutam Saha.
\newblock Blockchain-based security aspects in heterogeneous internet-of-things
  networks: A survey.
\newblock {\em Transactions on Emerging Telecommunications Technologies},
  30(10):e3741, 2019.

\bibitem{yang2019survey}
Wenli Yang, Erfan Aghasian, Saurabh Garg, David Herbert, Leandro Disiuta, and
  Byeong Kang.
\newblock A survey on blockchain-based internet service architecture:
  Requirements, challenges, trends and future.
\newblock {\em IEEE Access}, 2019.

\bibitem{salman2018security}
Tara Salman, Maede Zolanvari, Aiman Erbad, Raj Jain, and Mohammed Samaka.
\newblock Security services using blockchains: A state of the art survey.
\newblock {\em IEEE Communications Surveys \& Tutorials}, 21(1):858--880, 2018.

\bibitem{christidis2016blockchains}
Konstantinos Christidis and Michael Devetsikiotis.
\newblock Blockchains and smart contracts for the internet of things.
\newblock {\em IEEE Access}, 4:2292--2303, 2016.

\bibitem{atlam2018blockchain}
Hany~F Atlam, Ahmed Alenezi, Madini~O Alassafi, and Gary Wills.
\newblock Blockchain with internet of things: Benefits, challenges, and future
  directions.
\newblock {\em International Journal of Intelligent Systems and Applications},
  10(6):40--48, 2018.

\bibitem{xie2019survey}
Junfeng Xie, Helen Tang, Tao Huang, F~Richard Yu, Renchao Xie, Jiang Liu, and
  Yunjie Liu.
\newblock A survey of blockchain technology applied to smart cities: Research
  issues and challenges.
\newblock {\em IEEE Communications Surveys \& Tutorials}, 2019.

\bibitem{al2019blockchain}
Jameela Al-Jaroodi and Nader Mohamed.
\newblock Blockchain in industries: A survey.
\newblock {\em IEEE Access}, 7:36500--36515, 2019.

\bibitem{siano2019survey}
Pierluigi Siano, Giuseppe De~Marco, Alejandro Rol{\'a}n, and Vincenzo Loia.
\newblock A survey and evaluation of the potentials of distributed ledger
  technology for peer-to-peer transactive energy exchanges in local energy
  markets.
\newblock {\em IEEE Systems Journal}, 2019.

\bibitem{Lei_NGBN}
Zhibin Lei, Chao Feng, Yang Liu, Dennis~SF Lee, Tony Tsang, Jun Liang, Zhijun
  Xiong, Yuquan Liu, and Gang Chen.
\newblock Next generation blockchain network (ngbn).
\newblock In {\em 2019 20th IEEE International Conference on Mobile Data
  Management (MDM)}, pages 452--456. IEEE, 2019.

\bibitem{jiasi2019secure}
Jiasi Weng, Jian Weng, Jia-Nan Liu, and Yue Zhang.
\newblock Secure software-defined networking based on blockchain, 2019.

\bibitem{ripple}
David Schwartz, Noah Youngs, and Arthur Britto.
\newblock The ripple protocol consensus algorithm, 2014.

\bibitem{Yang_BNOPN}
Hui Yang, Yajie Li, Shaoyong Guo, Jian Ding, Young Lee, and Jie Zhang.
\newblock Distributed blockchain-based trusted control with multi-controller
  collaboration for software defined data center optical networks in 5g and
  beyond.
\newblock In {\em Optical Fiber Communication Conference}, pages Th1G--2.
  Optical Society of America, 2019.

\bibitem{Sharma_BCA}
Pradip~Kumar Sharma, Mu-Yen Chen, and Jong~Hyuk Park.
\newblock A software defined fog node based distributed blockchain cloud
  architecture for iot.
\newblock {\em IEEE Access}, 6:115--124, 2017.

\bibitem{Li_ICIOT}
Cheng Li and Liang-Jie Zhang.
\newblock A blockchain based new secure multi-layer network model for internet
  of things.
\newblock In {\em 2017 IEEE International Congress on Internet of Things
  (ICIOT)}, pages 33--41. IEEE, 2017.

\bibitem{Zhang_BNVANET}
Xiaodong Zhang, Ru~Li, and Bo~Cui.
\newblock A security architecture of vanet based on blockchain and mobile edge
  computing.
\newblock In {\em Proc. IEEE HotICN'18}, pages 258--259, Aug 2018.

\bibitem{III4}
Yong Yu, Yannan Li, Junfeng Tian, and Jianwei Liu.
\newblock Blockchain-based solutions to security and privacy issues in the
  internet of things.
\newblock {\em IEEE Wireless Communications}, 25(6):12--18, Dec. 2018.

\bibitem{III8}
Keke Gai, Yulu Wu, Liehuang Zhu, Lei Xu, and Yan Zhang.
\newblock Permissioned blockchain and edge computing empowered
  privacy-preserving smart grid networks.
\newblock {\em IEEE Internet of Things Journal}, pages 1--1, Mar. 2019.

\bibitem{Kobsa2007Privacy}
Alfred Kobsa.
\newblock Privacy-enhanced personalization.
\newblock {\em Commun. ACM}, 50(8):24–33, August 2007.

\bibitem{Hua2019CINEMA}
Jiafeng Hua, Hui Zhu, Fengwei Wang, Ximeng Liu, Rongxing Lu, Hao Li, and Yeping
  Zhang.
\newblock Cinema: Efficient and privacy-preserving online medical primary
  diagnosis with skyline query.
\newblock {\em IEEE Internet of Things Journal}, 6(2):1450--1461, 2019.

\bibitem{III11}
Pradip~Kumar Sharma, Saurabh Singh, Young-Sik Jeong, and Jong~Hyuk Park.
\newblock Distblocknet: A distributed blockchains-based secure sdn architecture
  for iot networks.
\newblock {\em IEEE Communications Magazine}, 55(9):78--85, Sep. 2017.

\bibitem{III15}
Libo Feng, Hui Zhang, Liqi Lou, and Yong Chen.
\newblock A blockchain-based collocation storage architecture for data security
  process platform of wsn.
\newblock {\em 2018 IEEE 22nd International Conference on Computer Supported
  Cooperative Work in Design}, pages 75--80, May. 2018.

\bibitem{Xu2019Efficient}
Yuhua Xu, Houtao Sun, Feng Xiang, and Zhixin Sun.
\newblock Efficient ddos detection based on k-fknn in software defined
  networks.
\newblock {\em IEEE Access}, 7:160536--160545, 2019.

\bibitem{He2020DNS}
Xudong He, Jian Wang, Jiqiang Liu, Zhen Han, Zhuo Lv, and Wei Wang.
\newblock Dns rebindingdetection for local internet of things devices.
\newblock In {\em Frontiers in Cyber Security}, pages 19--29, Singapore, 2020.
  Springer Singapore.

\bibitem{III16}
Hui Yang, Haowei Zheng, Jie Zhang, Yizhen Wu, Young Lee, and Yuefeng Ji.
\newblock Blockchain-based trusted authentication in cloud radio over fiber
  network for 5g.
\newblock {\em 16th International Conference on Optical Communications and
  Networks (ICOCN)}, pages 1--3, Aug. 2017.

\bibitem{III20}
Beini Zhou, Hui Li, and Li~Xu.
\newblock An authentication scheme using identity-based encryption blockchain.
\newblock {\em IEEE Symposium on Computers and Communications (ISCC)}, pages
  556--561, Jun. 2018.

\bibitem{Ma2010Privacy}
Chris~Y.T. Ma, David~K.Y. Yau, Nung~Kwan Yip, and Nageswara~S.V. Rao.
\newblock Privacy vulnerability of published anonymous mobility traces.
\newblock In {\em Proceedings of the Sixteenth Annual International Conference
  on Mobile Computing and Networking}, MobiCom '10, page 185–196, New York,
  NY, USA, 2010. Association for Computing Machinery.

\bibitem{Lu2012Pseudonym}
Rongxing Lu, Xiaodong Lin, Tom~H. Luan, Xiaohui Liang, and Xuemin Shen.
\newblock Pseudonym changing at social spots: An effective strategy for
  location privacy in vanets.
\newblock {\em IEEE Transactions on Vehicular Technology}, 61(1):86--96, 2012.

\bibitem{Wohlmacher2000Digital}
Petra Wohlmacher.
\newblock Digital certificates: A survey of revocation methods.
\newblock In {\em Proceedings of the 2000 ACM Workshops on Multimedia},
  MULTIMEDIA '00, page 111–114, New York, NY, USA, 2000. Association for
  Computing Machinery.

\bibitem{III23}
Shixiong Yao, Jing Chen, Kun He, Ruiying Du, Tianqing Zhu, and Xin Chen.
\newblock Pbcert: Privacy-preserving blockchain-based certificate status
  validation toward mass storage management.
\newblock {\em IEEE Access}, 7:6117--6128, Dec. 2019.

\bibitem{III24}
Xueping Liang, Sachin Shetty, Deepak Tosh, Charles Kamhoua, Kevin Kwiat, and
  Laurent Njilla.
\newblock Provchain: A blockchain-based data provenance architecture in cloud
  environment with enhanced privacy and availability.
\newblock In {\em Proceedings of the 17th IEEE/ACM international symposium on
  cluster, cloud and grid computing}, pages 468--477. IEEE Press, 2017.

\bibitem{yuan2018blockchain}
Yong Yuan and Fei-Yue Wang.
\newblock Blockchain and cryptocurrencies: Model, techniques, and applications.
\newblock {\em IEEE Transactions on Systems, Man, and Cybernetics: Systems},
  48(9):1421--1428, 2018.

\bibitem{wu2019comprehensive}
Mingli Wu, Kun Wang, Xiaoqin Cai, Song Guo, Minyi Guo, and Chunming Rong.
\newblock A comprehensive survey of blockchain: From theory to iot applications
  and beyond.
\newblock {\em IEEE Internet of Things Journal}, 2019.

\bibitem{watanabe2015blockchain}
Hiroki Watanabe, Shigeru Fujimura, Atsushi Nakadaira, Yasuhiko Miyazaki,
  Akihito Akutsu, and Jay~Junichi Kishigami.
\newblock Blockchain contract: A complete consensus using blockchain.
\newblock In {\em 2015 IEEE 4th global conference on consumer electronics
  (GCCE)}, pages 577--578. IEEE, 2015.

\bibitem{Zhang2019ACM}
Rui Zhang, Rui Xue, and Ling Liu.
\newblock Security and privacy on blockchain.
\newblock {\em ACM Comput. Surv.}, 52(3):1--34, July 2019.

\bibitem{neudecker2018network}
Till Neudecker and Hannes Hartenstein.
\newblock Network layer aspects of permissionless blockchains.
\newblock {\em IEEE Communications Surveys \& Tutorials}, 21(1):838--857, 2018.

\bibitem{wang2019survey}
Wenbo Wang, Dinh~Thai Hoang, Peizhao Hu, Zehui Xiong, Dusit Niyato, Ping Wang,
  Yonggang Wen, and Dong~In Kim.
\newblock A survey on consensus mechanisms and mining strategy management in
  blockchain networks.
\newblock {\em IEEE Access}, 7:22328--22370, 2019.

\bibitem{belotti2019vademecum}
Marianna Belotti, Nikola Bo{\v{z}}i{\'c}, Guy Pujolle, and Stefano Secci.
\newblock A vademecum on blockchain technologies: When, which and how.
\newblock {\em IEEE Communications Surveys \& Tutorials}, 2019.

\bibitem{Lu_BNFL}
Yunlong Lu, Xiaohong Huang, Yueyue Dai, Sabita Maharjan, and Yan Zhang.
\newblock Blockchain and federated learning for privacy-preserved data sharing
  in industrial iot.
\newblock {\em IEEE Transactions on Industrial Informatics}, 2019.

\bibitem{Liu_BNDS_DRL}
Chi~Harold Liu, Qiuxia Lin, and Shilin Wen.
\newblock Blockchain-enabled data collection and sharing for industrial iot
  with deep reinforcement learning.
\newblock {\em IEEE Transactions on Industrial Informatics}, 2018.

\bibitem{Chen_DS_BN}
Wuhui Chen, Yufei Chen, Xu~Chen, and Zibin Zheng.
\newblock Toward secure data sharing for the iov: A quality-driven incentive
  mechanism with on-chain and off-chain guarantees.
\newblock {\em IEEE Internet of Things Journal}, 2019.

\bibitem{IVC23}
Shangping Wang, Yinglong Zhang, and Yaling Zhang.
\newblock A blockchain-based framework for data sharing with fine-grained
  access control in decentralized storage systems.
\newblock {\em IEEE Access}, 6:38437--38450, 2018.

\bibitem{IVC21}
Masayuki Fukumitsu, Shingo Hasegawa, Junya Iwazaki, Masao Sakai, and Daiki
  Takahashi.
\newblock A proposal of a secure p2p-type storage scheme by using the secret
  sharing and the blockchain.
\newblock In {\em 2017 IEEE 31st International Conference on Advanced
  Information Networking and Applications (AINA)}, pages 803--810. IEEE, 2017.

\bibitem{IVC22}
Dagang Li, Rong Du, Yue Fu, and Man~Ho Au.
\newblock Meta-key: A secure data-sharing protocol under blockchain-based
  decentralized storage architecture.
\newblock {\em IEEE Networking Letters}, 1(1):30--33, Mar. 2019.

\bibitem{IVC24}
Ravi~Kiran Raman and Lav~R Varshney.
\newblock Distributed storage meets secret sharing on the blockchain.
\newblock In {\em 2018 Information Theory and Applications Workshop (ITA)},
  pages 1--6. IEEE, 2018.

\bibitem{IVC25}
Xiaochen Zheng, Raghava~Rao Mukkamala, Ravi Vatrapu, and Joaqun Ordieres-Mere.
\newblock Blockchain-based personal health data sharing system using cloud
  storage.
\newblock In {\em 2018 IEEE 20th International Conference on e-Health
  Networking, Applications and Services (Healthcom)}, pages 1--6. IEEE, 2018.

\bibitem{Jin_BNMDS}
Hao Jin, Yan Luo, Peilong Li, and Jomol Mathew.
\newblock A review of secure and privacy-preserving medical data sharing.
\newblock {\em IEEE Access}, 7:61656--61669, 2019.

\bibitem{jiang2018BlocHIE}
Shan Jiang, Jiannong Cao, Hanqing Wu, Yanni Yang, Mingyu Ma, and Jianfei He.
\newblock Blochie: A blockchain-based platform for healthcare information
  exchange.
\newblock In {\em 2018 IEEE International Conference on Smart Computing
  (SMARTCOMP)}, pages 49--56. IEEE, 2018.

\bibitem{Xia_MED}
QI~Xia, Emmanuel~Boateng Sifah, Kwame~Omono Asamoah, Jianbin Gao, Xiaojiang Du,
  and Mohsen Guizani.
\newblock Medshare: Trust-less medical data sharing among cloud service
  providers via blockchain.
\newblock {\em IEEE Access}, 5:14757--14767, 2017.

\bibitem{zyskind2015decentralizing}
Guy Zyskind, Oz~Nathan, et~al.
\newblock Decentralizing privacy: Using blockchain to protect personal data.
\newblock In {\em 2015 IEEE Security and Privacy Workshops}, pages 180--184.
  IEEE, 2015.

\bibitem{loukil2018towards}
Faiza Loukil, Chirine Ghedira-Guegan, Khouloud Boukadi, and A{\"\i}cha~Nabila
  Benharkat.
\newblock Towards an end-to-end iot data privacy-preserving framework using
  blockchain technology.
\newblock In {\em International Conference on Web Information Systems
  Engineering}, pages 68--78. Springer, 2018.

\bibitem{kaaniche2017blockchain}
Nesrine Kaaniche and Maryline Laurent.
\newblock A blockchain-based data usage auditing architecture with enhanced
  privacy and availability.
\newblock In {\em 2017 IEEE 16th International Symposium on Network Computing
  and Applications (NCA)}, pages 1--5. IEEE, 2017.

\bibitem{chen2019deplest}
Yun Chen, Hui Xie, Kun Lv, Shengjun Wei, and Changzhen Hu.
\newblock Deplest: A blockchain-based privacy-preserving distributed database
  toward user behaviors in social networks.
\newblock {\em Information Sciences}, 2019.

\bibitem{vishwa2018blockchain}
Alka Vishwa and Farookh~Khadeer Hussain.
\newblock A blockchain based approach for multimedia privacy protection and
  provenance.
\newblock In {\em 2018 IEEE Symposium Series on Computational Intelligence
  (SSCI)}, pages 1941--1945. IEEE, 2018.

\bibitem{jiang2019Privacy}
Shan Jiang, Jiannong Cao, Julie~A. McCann, Yanni Yang, Yang Liu, Xiaoqing Wang,
  and Yuming Deng.
\newblock Privacy-preserving and efficient multi-keyword search over encrypted
  data on blockchain.
\newblock In {\em 2019 IEEE International Conference on Blockchain
  (Blockchain)}, pages 405--410. IEEE, 2019.

\bibitem{al2019privacy}
Abdullah Al~Omar, Md~Zakirul~Alam Bhuiyan, Anirban Basu, Shinsaku Kiyomoto, and
  Mohammad~Shahriar Rahman.
\newblock Privacy-friendly platform for healthcare data in cloud based on
  blockchain environment.
\newblock {\em Future Generation Computer Systems}, 95:511--521, 2019.

\bibitem{dagher2018ancile}
Gaby~G Dagher, Jordan Mohler, Matea Milojkovic, and Praneeth~Babu Marella.
\newblock Ancile: Privacy-preserving framework for access control and
  interoperability of electronic health records using blockchain technology.
\newblock {\em Sustainable Cities and Society}, 39:283--297, 2018.

\bibitem{magyar2017blockchain}
G{\'a}bor Magyar.
\newblock Blockchain: Solving the privacy and research availability tradeoff
  for ehr data: A new disruptive technology in health data management.
\newblock In {\em 2017 IEEE 30th Neumann Colloquium (NC)}, pages
  000135--000140. IEEE, 2017.

\bibitem{nortey2019privacy}
Richard~Nuetey Nortey, Li~Yue, Promise~Ricardo Agdedanu, and Michael Adjeisah.
\newblock Privacy module for distributed electronic health records (ehrs) using
  the blockchain.
\newblock In {\em 2019 IEEE 4th International Conference on Big Data Analytics
  (ICBDA)}, pages 369--374. IEEE, 2019.

\bibitem{wilkinson2014storj}
Shawn Wilkinson, Tome Boshevski, Josh Brandoff, and Vitalik Buterin.
\newblock Storj a peer-to-peer cloud storage network.
\newblock 2014.

\bibitem{do2017blockchain}
Hoang~Giang Do and Wee~Keong Ng.
\newblock Blockchain-based system for secure data storage with private keyword
  search.
\newblock In {\em 2017 IEEE World Congress on Services (SERVICES)}, pages
  90--93. IEEE, 2017.

\bibitem{ruj2018blockstore}
Sushmita Ruj, Mohammad~Shahriar Rahman, Anirban Basu, and Shinsaku Kiyomoto.
\newblock Blockstore: A secure decentralized storage framework on blockchain.
\newblock In {\em 2018 IEEE 32nd International Conference on Advanced
  Information Networking and Applications (AINA)}, pages 1096--1103. IEEE,
  2018.

\bibitem{li2018block}
Jiaxing Li, Jigang Wu, and Long Chen.
\newblock Block-secure: Blockchain based scheme for secure p2p cloud storage.
\newblock {\em Information Sciences}, 465:219--231, 2018.

\bibitem{zhao2018mchain}
Bo~Zhao, Peiru Fan, and Mingtao Ni.
\newblock Mchain: a blockchain-based vm measurements secure storage approach in
  iaas cloud with enhanced integrity and controllability.
\newblock {\em IEEE Access}, 6:43758--43769, 2018.

\bibitem{han2018architecture}
Huirui Han, Mengxing Huang, Yu~Zhang, and Uzair~Aslam Bhatti.
\newblock An architecture of secure health information storage system based on
  blockchain technology.
\newblock In {\em International Conference on Cloud Computing and Security},
  pages 578--588. Springer, 2018.

\bibitem{liu2018data}
Jingqiang Liu, Bin Li, Lizhang Chen, Meng Hou, Feiran Xiang, and Peijun Wang.
\newblock A data storage method based on blockchain for decentralization dns.
\newblock In {\em 2018 IEEE Third International Conference on Data Science in
  Cyberspace (DSC)}, pages 189--196. IEEE, 2018.

\bibitem{kumar2018secure}
Manish Kumar, Ashish~Kumar Singh, and TV~Suresh Kumar.
\newblock Secure log storage using blockchain and cloud infrastructure.
\newblock In {\em 2018 9th International Conference on Computing, Communication
  and Networking Technologies (ICCCNT)}, pages 1--4. IEEE, 2018.

\bibitem{Huang_TLBN}
Ke~Huang, Xiaosong Zhang, Yi~Mu, Fatemeh Rezaeibagha, Xiaojiang Du, and Nadra
  Guizani.
\newblock Achieving intelligent trust-layer for iot via self-redactable
  blockchain.
\newblock {\em IEEE Transactions on Industrial Informatics}, 2019.

\bibitem{Ma_TDM}
Zhaofeng Ma, Xiaochang Wang, Jain Deepak~Kumar, Khan Hanees, Zhen Wang, and
  Hongmin Gao.
\newblock A blockchain-based trusted data management scheme in edge computing.
\newblock {\em IEEE Transactions on Industrial Informatics}, 2019.

\bibitem{Zhang_BTM}
Yongping Zhang, Xiwei Xu, Ang Liu, Qinghua Lu, Lida Xu, and Fei Tao.
\newblock Blockchain-based trust mechanism for iot-based smart manufacturing
  system.
\newblock {\em IEEE Transactions on Computational Social Systems}, 2019.

\bibitem{Yang_TM_VN}
Zhe Yang, Kan Yang, Lei Lei, Kan Zheng, and Victor~CM Leung.
\newblock Blockchain-based decentralized trust management in vehicular
  networks.
\newblock {\em IEEE Internet of Things Journal}, 6(2):1495--1505, 2018.

\bibitem{III5}
Zhaojun Lu, Wenchao Liu, Qian Wang, Gang Qu, and Zhenglin Liu.
\newblock A privacy-preserving trust model based on blockchain for vanets.
\newblock {\em IEEE Access}, 6:45655--45664, Aug. 2018.

\bibitem{Yang_BTEV}
Yao-Tsung Yang, Li-Der Chou, Chia-Wei Tseng, Fan-Hsun Tseng, and Chien-Chang
  Liu.
\newblock Blockchain-based traffic event validation and trust verification for
  vanets.
\newblock {\em IEEE Access}, 7:30868--30877, 2019.

\bibitem{Ugobame}
Uchi~Ugobame Uchibeke, Kevin~A Schneider, Sara~Hosseinzadeh Kassani, and Ralph
  Deters.
\newblock Blockchain access control ecosystem for big data security.
\newblock In {\em 2018 IEEE International Conference on Internet of Things
  (iThings) and IEEE Green Computing and Communications (GreenCom) and IEEE
  Cyber, Physical and Social Computing (CPSCom) and IEEE Smart Data
  (SmartData)}, pages 1373--1378. IEEE, 2018.

\bibitem{Wang_ACBC}
Shangping Wang, Xu~Wang, and Yaling Zhang.
\newblock A secure cloud storage framework with access control based on
  blockchain.
\newblock {\em IEEE Access}, 7:112713--112725, 2019.

\bibitem{Tan_ACBC}
Xiaobin Tan, Chaoming Huang, and Liguo Ji.
\newblock Access control scheme based on combination of blockchain and
  xor-coding for icn.
\newblock In {\em 2018 5th IEEE International Conference on Cyber Security and
  Cloud Computing (CSCloud)/2018 4th IEEE International Conference on Edge
  Computing and Scalable Cloud (EdgeCom)}, pages 160--165. IEEE, 2018.

\bibitem{Ding_ABAC}
Sheng Ding, Jin Cao, Chen Li, Kai Fan, and Hui Li.
\newblock A novel attribute-based access control scheme using blockchain for
  iot.
\newblock {\em IEEE Access}, 7:38431--38441, 2019.

\bibitem{Laurent}
Sophie Dram{\'e}-Maign{\'e}, Maryline Laurent, and Laurent Castillo.
\newblock Distributed access control solution for the iot based on
  multi-endorsed attributes and smart contracts.
\newblock In {\em 2019 15th International Wireless Communications \& Mobile
  Computing Conference (IWCMC)}, pages 1582--1587. IEEE, 2019.

\bibitem{Ma_ACBC}
Mingxin Ma, Guozhen Shi, and Fenghua Li.
\newblock Privacy-oriented blockchain-based distributed key management
  architecture for hierarchical access control in the iot scenario.
\newblock {\em IEEE Access}, 7:34045--34059, 2019.

\bibitem{Paillisse}
Jordi Paillisse, Jordi Subira, Albert Lopez{-}Bresco, Alberto
  Rodr{\'{\i}}guez{-}Natal, Vina Ermagan, Fabio Maino, and Albert Cabellos.
\newblock Distributed access control with blockchain.
\newblock In {\em ICC 2019 - 2019 IEEE International Conference on
  Communications (ICC)}, pages 1--6, May 2019.

\bibitem{meng2018intrusion}
Weizhi Meng, Elmar~Wolfgang Tischhauser, Qingju Wang, Yu~Wang, and Jinguang
  Han.
\newblock When intrusion detection meets blockchain technology: A review.
\newblock {\em IEEE Access}, 6:10179--10188, 2018.

\bibitem{ujjan2019snort}
Raja Majid~Ali Ujjan, Zeeshan Pervez, and Keshav Dahal.
\newblock Snort based collaborative intrusion detection system using blockchain
  in sdn.
\newblock In {\em 2019 13th International Conference on Software, Knowledge,
  Information Management and Applications (SKIMA)}, pages 1--8. IEEE, 2019.

\bibitem{alkadi2020deep}
Osama Alkadi, Nour Moustafa, Benjamin Turnbull, and Kim-Kwang~Raymond Choo.
\newblock A deep blockchain framework-enabled collaborative intrusion detection
  for protecting iot and cloud networks.
\newblock {\em IEEE Internet of Things Journal}, 2020.

\bibitem{Saad_Routing}
Muhammad Saad, Afsah Anwar, Ashar Ahmad, Hisham Alasmary, Murat Yuksel, and
  Aziz Mohaisen.
\newblock Routechain: Towards blockchain-based secure and efficient bgp
  routing.
\newblock In {\em 2019 IEEE International Conference on Blockchain and
  Cryptocurrency (ICBC)}, pages 210--218. IEEE, 2019.

\bibitem{li2019mitigating}
Min Li, Helen Tang, and Xianbin Wang.
\newblock Mitigating routing misbehavior using blockchain-based distributed
  reputation management system for iot networks.
\newblock In {\em 2019 IEEE International Conference on Communications
  Workshops (ICC Workshops)}, pages 1--6. IEEE, 2019.

\bibitem{ramezan2018blockchain}
Gholamreza Ramezan and Cyril Leung.
\newblock A blockchain-based contractual routing protocol for the internet of
  things using smart contracts.
\newblock {\em Wireless Communications and Mobile Computing}, 2018, 2018.

\bibitem{yang2019trusted}
Jidian Yang, Shiwen He, Yang Xu, Linweiya Chen, and Ju~Ren.
\newblock A trusted routing scheme using blockchain and reinforcement learning
  for wireless sensor networks.
\newblock {\em Sensors}, 19(4):970, 2019.

\bibitem{dinginterchain}
Donghui Ding, Tiantian Duan, Linpeng Jia, Kang Li, Zhongcheng Li, and Yi~Sun.
\newblock Interchain: A framework to support blockchain interoperability.
\newblock 2018.

\bibitem{chen2017inter}
Zhi-dong Chen, YU~Zhuo, Zhang-bo Duan, and HU~Kai.
\newblock Inter-blockchain communication.
\newblock {\em DEStech Transactions on Computer Science and Engineering},
  (cst), 2017.

\bibitem{Anlink}
ZhongAn Tech.
\newblock Anlink blockchain network whitepaper, 2017.

\bibitem{ersoy2018transaction}
O{\u{g}}uzhan Ersoy, Zhijie Ren, Zekeriya Erkin, and Reginald~L Lagendijk.
\newblock Transaction propagation on permissionless blockchains: Incentive and
  routing mechanisms.
\newblock In {\em 2018 Crypto Valley Conference on Blockchain Technology
  (CVCBT)}, pages 20--30. IEEE, 2018.

\bibitem{rawat2019fusion}
Danda~B Rawat.
\newblock Fusion of software defined networking, edge computing, and blockchain
  technology for wireless network virtualization.
\newblock {\em IEEE Communications Magazine}, 57(10):50--55, 2019.

\bibitem{samaniego2017virtual}
Mayra Samaniego and Ralph Deters.
\newblock Virtual resources \& blockchain for configuration management in iot.
\newblock {\em Journal of Ubiquitous Systems \& Pervasive Networks},
  9(2):01--13, 2017.

\bibitem{zhu2018edgechain}
He~Zhu, Changcheng Huang, and Jiayu Zhou.
\newblock Edgechain: Blockchain-based multi-vendor mobile edge application
  placement.
\newblock In {\em 2018 4th IEEE Conference on Network Softwarization and
  Workshops (NetSoft)}, pages 222--226. IEEE, 2018.

\bibitem{backman2017blockchain}
Jere Backman, Seppo Yrj{\"o}l{\"a}, Kristiina Valtanen, and Olli
  M{\"a}mmel{\"a}.
\newblock Blockchain network slice broker in 5g: Slice leasing in factory of
  the future use case.
\newblock In {\em 2017 Internet of Things Business Models, Users, and
  Networks}, pages 1--8. IEEE, 2017.

\bibitem{li2019blockchain}
Ming Li and GQ~Huang.
\newblock Blockchain-enabled workflow management system for fine-grained
  resource sharing in e-commerce logistics.
\newblock In {\em 2019 IEEE 15th International Conference on Automation Science
  and Engineering (CASE)}, pages 751--755. IEEE, 2019.

\bibitem{zhou2018beekeeper}
Lijing Zhou, Licheng Wang, Yiru Sun, and Pin Lv.
\newblock Beekeeper: A blockchain-based iot system with secure storage and
  homomorphic computation.
\newblock {\em IEEE Access}, 6:43472--43488, 2018.

\bibitem{shi2018mpcstoken}
Fengrui Shi, Zhijin Qin, Di~Wu, and Julie McCann.
\newblock Mpcstoken: Smart contract enabled fault-tolerant incentivisation for
  mobile p2p crowd services.
\newblock In {\em 2018 IEEE 38th International Conference on Distributed
  Computing Systems (ICDCS)}, pages 961--971. IEEE, 2018.

\bibitem{sagirlar2018autobotcatcher}
Gokhan Sagirlar, Barbara Carminati, and Elena Ferrari.
\newblock Autobotcatcher: Blockchain-based p2p botnet detection for the
  internet of things.
\newblock In {\em 2018 IEEE 4th International Conference on Collaboration and
  Internet Computing (CIC)}, pages 1--8. IEEE, 2018.

\bibitem{mylrea2017blockchain}
Michael Mylrea and Sri Nikhil~Gupta Gourisetti.
\newblock Blockchain for smart grid resilience: Exchanging distributed energy
  at speed, scale and security.
\newblock In {\em 2017 Resilience Week (RWS)}, pages 18--23. IEEE, 2017.

\bibitem{liang2017towards}
Xueping Liang, Juan Zhao, Sachin Shetty, and Danyi Li.
\newblock Towards data assurance and resilience in iot using blockchain.
\newblock In {\em MILCOM 2017-2017 IEEE Military Communications Conference
  (MILCOM)}, pages 261--266. IEEE, 2017.

\bibitem{cachin2016architecture}
Christian Cachin.
\newblock Architecture of the hyperledger blockchain fabric.
\newblock In {\em Workshop on distributed cryptocurrencies and consensus
  ledgers}, volume 310, page~4, 2016.

\bibitem{eyal2018majority}
Ittay Eyal and Emin~G{\"u}n Sirer.
\newblock Majority is not enough: Bitcoin mining is vulnerable.
\newblock {\em Communications of the ACM}, 61(7):95--102, 2018.

\bibitem{natoli2016balance}
Christopher Natoli and Vincent Gramoli.
\newblock The balance attack against proof-of-work blockchains: The r3 testbed
  as an example.
\newblock {\em arXiv preprint arXiv:1612.09426}, 2016.

\bibitem{apostolaki2017hijacking}
Maria Apostolaki, Aviv Zohar, and Laurent Vanbever.
\newblock Hijacking bitcoin: Routing attacks on cryptocurrencies.
\newblock In {\em 2017 IEEE Symposium on Security and Privacy (SP)}, pages
  375--392. IEEE, 2017.

\bibitem{singh2006eclipse}
Atul Singh et~al.
\newblock Eclipse attacks on overlay networks: Threats and defenses.
\newblock In {\em In IEEE INFOCOM}. Citeseer, 2006.

\bibitem{Aggarwal2017Quantum}
Divesh Aggarwal, Gavin~K. Brennen, Troy Lee, Miklos Santha, and Marco
  Tomamichel.
\newblock Quantum attacks on bitcoin, and how to protect against them.
\newblock {\em Papers}, 2017.

\bibitem{yue2017big}
Yue Li, Junqin Huang, Shengzhi Qin, and Ruijin Wang.
\newblock Big data model of security sharing based on blockchain.
\newblock In {\em 2017 3rd International Conference on Big Data Computing and
  Communications (BIGCOM)}, pages 117--121. IEEE, 2017.

\bibitem{es2017blockchain}
Hamza Es-Samaali, Aissam Outchakoucht, and Jean~Philippe Leroy.
\newblock A blockchain-based access control for big data.
\newblock {\em International Journal of Computer Networks and Communications
  Security}, 5(7):137, 2017.

\bibitem{ShardingBlockchains}
Vitalik Buterin.
\newblock On sharding blockchains.
\newblock 2017.

\bibitem{miniblockchain}
Bruce~J D.
\newblock The mini-blockchain scheme.
\newblock {\em White paper}, 2014.

\bibitem{Perard2018Erasure}
Doriane Perard, Jerome Lacan, Yann Bachy, and Jonathan Detchart.
\newblock Erasure code-based low storage blockchain node.
\newblock In {\em 2018 IEEE International Conference on Internet of Things
  (iThings) and IEEE Green Computing and Communications (GreenCom) and IEEE
  Cyber, Physical and Social Computing (CPSCom) and IEEE Smart Data
  (SmartData)}, 2018.

\bibitem{Dorri2017Towards}
Ali Dorri, Salil~S Kanhere, and Raja Jurdak.
\newblock Towards an optimized blockchain for iot.
\newblock In {\em The second IEEE/ACM conference on Internet of Things Design
  and Implementation, IoTDI 2017}, 2017.

\bibitem{Plasma}
Buterin~V Poon~J.
\newblock Plasma: Scalable autonomous smart contracts.
\newblock {\em White paper}, pages 1--47, 2017.

\bibitem{Maksymyuk2019}
Taras Maksymyuk, Juraj Gazda, Longzhe Han, and Minho Jo.
\newblock Blockchain-based intelligent network management for 5g and beyond.
\newblock In {\em 2019 3rd International Conference on Advanced Information and
  Communications Technologies (AICT)}, pages 36--39, July 2019.

\bibitem{panikkar2015adept}
Sanjay Panikkar, Sumabala Nair, Paul Brody, and Veena Pureswaran.
\newblock Adept: An iot practitioner perspective.
\newblock {\em Draft Copy for Advance Review, IBM}, 2015.

\bibitem{chohan2017decentralized}
Usman~W Chohan.
\newblock The decentralized autonomous organization and governance issues.
\newblock {\em Available at SSRN 3082055}, 2017.

\bibitem{ali2017iot}
Muhammad~Salek Ali, Koustabh Dolui, and Fabio Antonelli.
\newblock Iot data privacy via blockchains and ipfs.
\newblock In {\em Proceedings of the Seventh International Conference on the
  Internet of Things}, page~14. ACM, 2017.

\bibitem{xu2017intelligent}
Chenhan Xu, Kun Wang, and Mingyi Guo.
\newblock Intelligent resource management in blockchain-based cloud
  datacenters.
\newblock {\em IEEE Cloud Computing}, 4(6):50--59, 2017.

\bibitem{ball2017proofs}
Marshall Ball, Alon Rosen, Manuel Sabin, and Prashant~Nalini Vasudevan.
\newblock Proofs of useful work.
\newblock {\em IACR Cryptology ePrint Archive}, 2017:203, 2017.

\bibitem{zhang2017rem}
Fan Zhang, Ittay Eyal, Robert Escriva, Ari Juels, and Robbert Van~Renesse.
\newblock $\{$REM$\}$: Resource-efficient mining for blockchains.
\newblock In {\em 26th $\{$USENIX$\}$ Security Symposium ($\{$USENIX$\}$
  Security 17)}, pages 1427--1444, 2017.

\bibitem{park2015spacecoin}
Sunoo Park, Krzysztof Pietrzak, Jo{\"e}l Alwen, Georg Fuchsbauer, and Peter
  Gazi.
\newblock Spacecoin: A cryptocurrency based on proofs of space.
\newblock In {\em IACR Cryptology ePrint Archive}. 2015.

\bibitem{xu2019become}
Xiaolong Xu, Xuyun Zhang, Honghao Gao, Yuan Xue, Lianyong Qi, and Wanchun Dou.
\newblock Become: Blockchain-enabled computation offloading for iot in mobile
  edge computing.
\newblock {\em IEEE Transactions on Industrial Informatics}, 2019.

\bibitem{luong2018optimal}
Nguyen~Cong Luong, Zehui Xiong, Ping Wang, and Dusit Niyato.
\newblock Optimal auction for edge computing resource management in mobile
  blockchain networks: A deep learning approach.
\newblock In {\em 2018 IEEE International Conference on Communications (ICC)},
  pages 1--6. IEEE, 2018.

\bibitem{lima2018developing}
Claudio Lima.
\newblock Developing open and interoperable dlt$\backslash$/blockchain
  standards [standards].
\newblock {\em Computer}, 51(11):106--111, 2018.

\end{thebibliography}

\end{document}